\documentclass[journal]{vgtc}

\onlineid{1512}

\vgtccategory{Research}
\vgtcpapertype{Technique}

\usepackage{tabu}
\usepackage{booktabs}
\usepackage{amsfonts}
\usepackage{amsmath}
\usepackage{multirow}

\newcommand{\N}{\mathbb{N}}
\newcommand{\R}{\mathbb{R}}
\newcommand{\Z}{\mathbb{Z}}
\newcommand{\X}{\mathbb{X}}

\newcommand{\para}[1]{\vspace{1pt}\noindent{\textbf{#1}}}

\title{A General Framework for Augmenting Lossy Compressors with Topological Guarantees}

\author{
  \authororcid{Nathaniel Gorski}{0009-0001-8205-5640}, \authororcid{Xin Liang}{0000-0002-0630-1600}, \authororcid{Hanqi Guo}{0000-0001-7776-1834},\authororcid{Lin Yan}{0000-0001-7017-0329}, and \authororcid{Bei Wang}{0000-0002-9240-0700}
}

\authorfooter{
\item Nathaniel Gorski is with the University of Utah.
  	E-mail: gorski@sci.utah.edu.
\item Xin Liang is with the University of Kentucky. 
 	E-mail: xliang@uky.edu.
\item Hanqi Guo is with the Ohio State University. 
	E-mail: guo.2154@osu.edu.
\item Lin Yan is with the Iowa State University. 
	E-mail: linyan@iastate.edu. 
\item Bei Wang is with the University of Utah. 
	E-mail: beiwang@sci.utah.edu. 
}

\abstract{
Topological descriptors such as contour trees are widely utilized in scientific data analysis and visualization, with applications from materials science to climate simulations. 
It is desirable to preserve topological descriptors when data compression is part of the scientific workflow for these applications. 
However, classic error-bounded lossy compressors for volumetric data do not guarantee the preservation of topological descriptors, despite imposing strict pointwise error bounds. 
In this work, we introduce a general framework for augmenting \emph{any} lossy compressor to preserve the topology of the data during compression. 
Specifically, our framework quantifies the adjustments (to the decompressed data) needed to preserve the contour tree and then employs a custom variable-precision encoding scheme to store these adjustments. 
We demonstrate the utility of our framework in augmenting classic compressors (such as SZ3, TTHRESH, and ZFP) and deep learning-based compressors (such as Neurcomp) with topological guarantees.
}

\keywords{Lossy compression, contour tree, topology preservation, topological data analysis, topology in visualization}

\teaser{
\centering
  \begin{subfigure}[b]{0.02\textwidth}
    \raisebox{\height}{\includegraphics[angle=90,width=\textwidth]{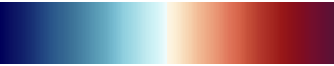}}
  \end{subfigure}
  \begin{subfigure}[b]{0.97\textwidth}
    \includegraphics[width=\textwidth]{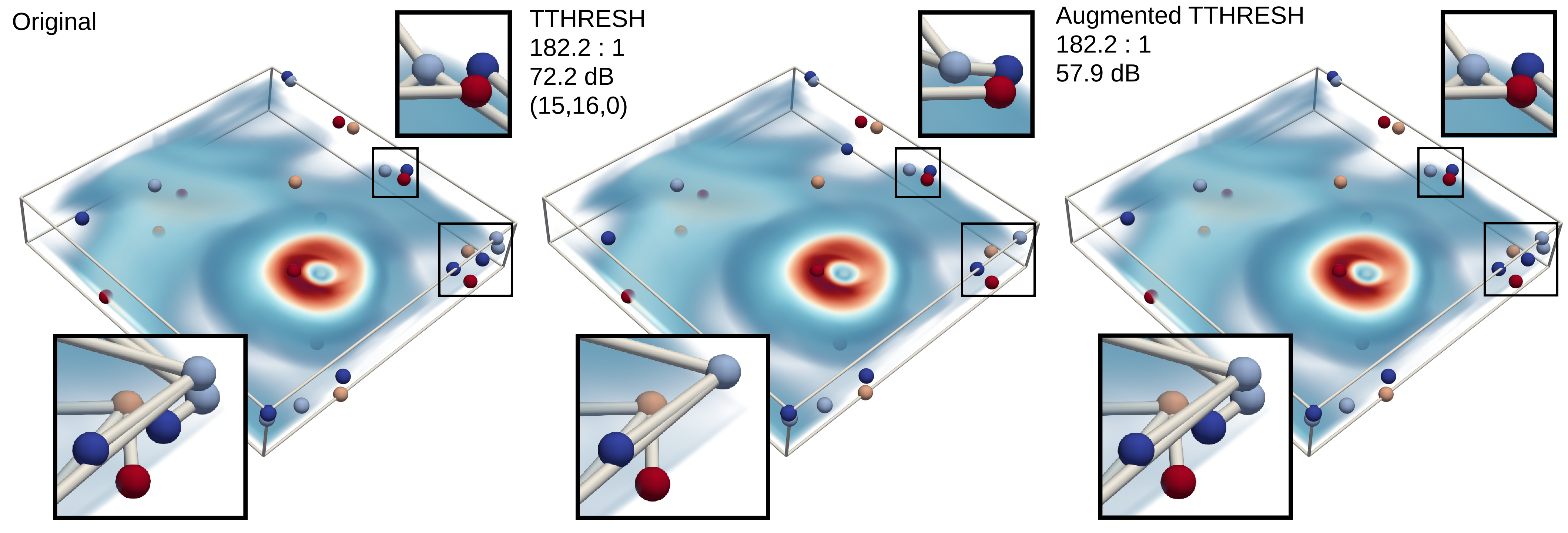}
  \end{subfigure}
\caption{Visualizing the Isabel dataset compressed with classic TTHRESH and Augmented TTHRESH compressors with topological controls. 
Left: input data visualized with critical points of the contour tree; maximum are in red, minimum are in dark blue, 1-saddles are in light blue, and 2-saddles are in light orange.  
Middle: reconstructed data using TTHRESH, labeled with compression ratio, PSNR, and false cases (false positives, false negative, false types).
Right: reconstructed data using Augmented TTHRESH along with compression ratio and PSNR. 
We also provide zoomed-in views that highlight the connectivity among critical points of the contour trees.}
\label{fig:teaser}
}

\begin{document}

\maketitle

%-------------------------
\section{Introduction}
\label{sec:introduction}

Modern scientific simulations generate enormous amounts of data, such that it is not practical to store all of the data produced \cite{cappello2019use}.
Data compression helps reduce the size of data, making data storage more practical.
%for storage and transmission in scientific data management systems. 
There are two types of compression techniques: lossy and lossless.
Lossy compression techniques allow for some distortion of the data to achieve smaller compressed file sizes and have wide applicability in the compression of images \cite{hussain2018image}, audio \cite{kavitha2016survey}, and scientific data \cite{cappello2019use}. 
Among them, error-bounded lossy compressors, such as SZ \cite{liang2022sz3}, ZFP \cite{lindstrom2014fixed}, and TTHRESH \cite{ballester2019tthresh}, play a crucial role in reducing the storage demand of large-scale scientific data. 
Not only can such compressors significantly reduce the data volume, but also they can control the data distortion and guarantee the validity of the reconstructed data for post hoc analysis, based on user-defined error bounds~\cite{LiuDiZhao2022}. 
For instance, error-bounded lossy compressors have been shown to reduce file size dramatically without significant degradation of the visual quality on the reconstructed (decompressed) data (e.g.,~\cite{LiangDiTao2019, ballester2019tthresh,zhao2021optimizing}). 

In the analysis of scientific data, topological data analysis (TDA) employs topological descriptors, such as contour trees~\cite{carr2003computing} and Morse--Smale complexes~\cite{edelsbrunner2001hierarchical}, to describe, summarize, and draw conclusions about scientific data (e.g.,\cite{aydogan2014characterization, cazals2003molecular,rosen2021using}); see~\cite{YanMasoodSridharamurthy2021} for a survey. 
Although error-bounded lossy compressors typically allow the user to impose pointwise error bounds that are maintained during compression, such compressors seldom make guarantees about how the compression will affect the geometry and topology of the reconstructed data. 

For example, Lu et al.~\cite{lu2018understanding} examined the impact of lossy compression on data fidelity and complex scientific data analytics.
They studied the detection of \emph{blobs}, features used by fusion scientists to study the trajectory of high-energy particles. 
Blobs are defined by areas with high electric potentials, i.e., areas enclosed by contours above certain thresholds. 
Their experiments demonstrated that as the error bound increases, the blobs will change in both number and position. 
They concluded that ``lossy compression may seriously distort
data, thus having a disastrous impact on data analytics,'' and therefore ``determining a proper error bound is key to performing
meaningful lossy compression in science production.''~\cite{lu2018understanding} 
We argue that determining a proper error bound is only part of the story. It is also important to develop error-bounded lossy compressors that explicitly preserve features of interest to domain scientists---such as topological features---during compression. 

In this paper, we introduce a general framework that augments \emph{any} lossy compressor for volumetric scalar fields---error bounded or not---in order to impose topological control, while maintaining a user-specified pointwise error bound.  
Specifically, our framework augments any lossy compressor in order to preserve the \emph{contour tree} in terms of its critical points and the connectivity among those critical points. 
Preserving the contour tree of the reconstructed data is crucial to support a variety of post hoc scientific visualization tasks, since the contour tree has been used for feature extraction, tracking, comparison, and interactive contour exploration (e.g.,~\cite{zhou2009automatic, kopp2022temporal}). 

Previous work by Yan et al.~\cite{yan2023toposz} modified the classic SZ compressor with a customized error-controlled quantization strategy to preserve the contour tree. 
Instead of modifying any \emph{single} compressor~\cite{yan2023toposz}, our general framework can augment \emph{any} scalar field compressor, with \emph{no restrictions} on the base compressor. We effectively leverage the capabilities of a wide variety of data compressors to preserve the contour tree. 
Additionally, as data compressors continue to improve, our framework can be used to augment increasingly effective compressors and thereby achieve better results. 
 
During augmentation, our framework introduces a progressive strategy to compute upper and lower error bounds for specific data points, which, if maintained, will guarantee that the contour tree is preserved and the pointwise error bound is maintained. 
The classic algorithm constructs a contour tree by combining two merge trees~\cite{carr2003computing} (i.e., a join tree and a split tree, see~\cref{sec:merge-and-contour-tree} for details).  
Our progressive strategy works with these merge trees: if the merge trees are preserved, then so is the contour tree.
Our framework is \emph{progressive} in the sense that it works through a merge tree computation, preserving one branch at a time. 
Only a small portion of the merge tree is computed more than once during the error bound tightening process.
This is in contrast with TopoSZ, which iteratively recomputes the entire contour tree many times. Thus, our progressive approach shows significant speedups compared with TopoSZ.
Our strategy then uses a novel variable-precision encoding scheme to store any adjustments that must be made to the output of an augmented compressor in order to ensure that these upper and lower bounds are maintained.  
Our contributions include: 
\begin{itemize}[noitemsep,leftmargin=*]
\item A novel progressive strategy that efficiently refines upper and lower bounds during the merge tree computations (subroutines for contour tree computation) without computing contour trees explicitly (as in TopoSZ). These error bounds correspond to adjustments that must be made to the output of any compressor in order to preserve the contour tree.   
\item A custom variable-precision encoding scheme to efficiently store these adjustments.
\item A systemic comparative study that evaluates five lossy compressors (ZFP, SZ3, Cubic Spline, TTHRESH, Neurcomp) augmented with topological control, and two state-of-the-art topology-preserving compressors, across a number of scientific datasets.  
\end{itemize}
Our experimental study demonstrates the effectiveness and efficiency of our framework in enabling topological controls for a wide variety of lossy compressors.
\section{Related Work}
\label{sec:related-work}

We give a brief review of data compression for volumetric data. 
We then discuss the use of contour trees in topological data analysis, followed by related work for topology-preserving compression techniques.   

\para{Lossy compression.} 
Lossless compression techniques allow the original data to be perfectly reconstructed, 
but they usually suffer from limited compression ratios (less than $2\times$ according to~\cite{son2014data}) in scientific data and thus are not practical. 
Lossy compression is an alternative way to reduce the unprecedented  size of scientific data. 
Traditional lossy techniques such as JPEG/JPEG2000 leverage wavelet theories and bit plane encoding to compress image data, but they are not adept at dealing with multidimensional scientific data in floating-point format. 
Recently, there has been an increasing trend to leverage deep learning techniques, such as the autoencoder~\cite{le2023hierarchical} and implicit neural representation (INR)~\cite{lu2021compressive}, for data compression.
An autoencoder is a neural network composed of two components: an encoder and a decoder. 
The encoder is trained to produce low-dimensional representations of the input data, whereas the decoder is trained to reconstruct the original input data from the output of the encoder. 
An INR model trains a small neural network that can be used to recreate the ground truth. 
The neural network itself is shipped as a compressed file, and to decompress it, one must simply evaluate the network on an appropriate input. 
One notable INR model for volumetric scalar fields is Neurcomp~\cite{lu2021compressive}.
Recently, spatial super-resolution (SSR) models have employed neural networks to accurately upscale low-resolution representation of data as a form of interpolation. 
Several volumetric scalar field compressors incorporate SSR models, such as SSR-TVD~\cite{han2020ssr} and the deep hierarchical model~\cite{wurster2022deep}.
Unfortunately, these general lossy techniques lack precise error control on the data, which limits their use on scientific data.

Error-controlled lossy compressors~\cite{lindstrom2014fixed,ballester2019tthresh,zhao2021optimizing,lakshminarasimhan2013isabela} have been proposed and leveraged by the scientific computing community to reduce the data size while controlling the distortion in the decompressed data. 
In general, such compressors can be categorized into transform-based and prediction-based. 
Transform-based lossy compressors rely on domain transforms for data decorrelation. 
For instance, ZFP~\cite{lindstrom2014fixed} divides data into small blocks and then compresses each block independently. The compression procedure inside each block includes exponent alignment for fixed point conversion, a near-orthogonal domain transform, and embedded encoding. 
TTHRESH~\cite{ballester2019tthresh} is another transform-based compressor that leverages singular value decomposition (SVD) to improve the decorrelation efficiency for high-dimensional data.

Prediction-based compressors employ prediction methods such as interpolation to approximate the ground truth. The differences between original and predicted data are quantized and then encoded using entropy encoding and lossless techniques. 
ISABELA~\cite{lakshminarasimhan2013isabela}, as one of the pioneering error-controlled prediction-based compressors, uses B-splines to predict data. 
SZ3~\cite{liang2022sz3,zhao2021optimizing,liang2018error}, the most recent general release in the SZ compressor family, uses a combination of a Lorenzo predictor~\cite{ibarria2003out}, cubic spline interpolation, and linear interpolation. 
In addition, AE-SZ~\cite{liu2021exploring} is proposed as a variation of SZ that incorporates autoencoders in the prediction pipeline.

\para{Contour trees.} Our augmented compressors aim to preserve the contour tree of an input scalar field. 
Contour trees capture the relationships among contours of scalar fields. 
They have been used to support data analysis and visualization tasks across diverse disciplines, such as astronomy \cite{rosen2021using}, fluid dynamics \cite{aydogan2014characterization}, and medicine \cite{aydogan2013analysis, wang2018fast, szymczak2010categorical}. 
They have also been incorporated into algorithms in computer vision \cite{mizuta2005description} and visualization \cite{zhou2009automatic, kopp2022temporal} for interactive exploration of contours.  

\para{Topology-preserving compression.} 
To the best of our knowledge, only three compressors have been developed for scalar field compression with topological guarantees. 
The first compressor was developed by Soler et al. \cite{soler2018topologically}. We shall refer to it as TopoQZ. 
TopoQZ allows the user to specify a single parameter $\varepsilon$. 
It preserves all critical point pairs with finite persistence greater than $\varepsilon$ and eliminates all critical points with persistence less than $\varepsilon$. 
TopoQZ is not designed to perfectly preserve the contour tree. Therefore, the locations of preserved critical points may shift slightly during compression, and the connectivity of the critical points in the contour tree may be altered. 
TopoQZ can also guarantee that the reconstructed values differ from the ground truth at most by a user-specified error bound $\xi$. It is required that $\xi > \varepsilon$.
TopoQZ is currently implemented in the Topology Toolkit \cite{TiernyFavelierLevine2017, MasoodBudinFalk2021, leguillou_tvcg24}. That implementation couples TopoQZ with ZFP \cite{lindstrom2014fixed}, which improves the smoothness of the data but introduces additional pointwise error.

Another topology-preserving compressor is TopoSZ \cite{yan2023toposz}. 
TopoSZ modifies the classic SZ pipeline to perfectly preserve the contour tree of the ground truth data up to the persistence threshold of $\varepsilon$. 
That is, the contour tree of the output of TopoSZ will be equal to that of the ground truth after both datasets have been topologically simplified with a persistence threshold of $\varepsilon$. 
TopoSZ also allows the user to impose a strict error-bound $\xi$ (and allows $\xi \leq \varepsilon$). When compared with TopoQZ, TopoSZ yields generally higher compression ratios and reconstruction quality, although the algorithm takes longer to execute. 
Our general framework borrows some elements from the TopoSZ pipeline. However, our framework differs significantly from TopoSZ due to two technical innovations: progressive bound tightening and logarithmic-scaling quantization (see \cref{sec:method} for details).  

Most recently, Li et al. developed mSZ~\cite{li2024msz} that augments an existing lossy compressor to compress a 2D/3D scalar field while preserving its piecewise-linear (PL) Morse--Smale segmentation \cite{edelsbrunner2001hierarchical,edelsbrunner2003morse},~i.e.,~a partition of the data domain based on the Morse--Smale complex. 
In comparison to the contour tree that is based on the level sets of a scalar field, a Morse--Smale complex is a different topological descriptor based on the gradient behavior of a scalar field.   
Because our framework instead preserves contour trees and does not consider the gradients in its pipeline, mSZ is not directly comparable to our work. 

Finally, even though it does not preserve any common topological descriptor, cpSZ \cite{liang2022toward}---a variation of SZ---preserves the critical points of a vector field. cpSZ also introduces a log-scale quantization technique to store different error bounds for individual points.

\section{Technical Background}
\label{sec:background}

\subsection{Merge Tree and Contour Tree}
\label{sec:merge-and-contour-tree}
\begin{figure}[!ht]
    \centering
    \includegraphics[width=\linewidth]{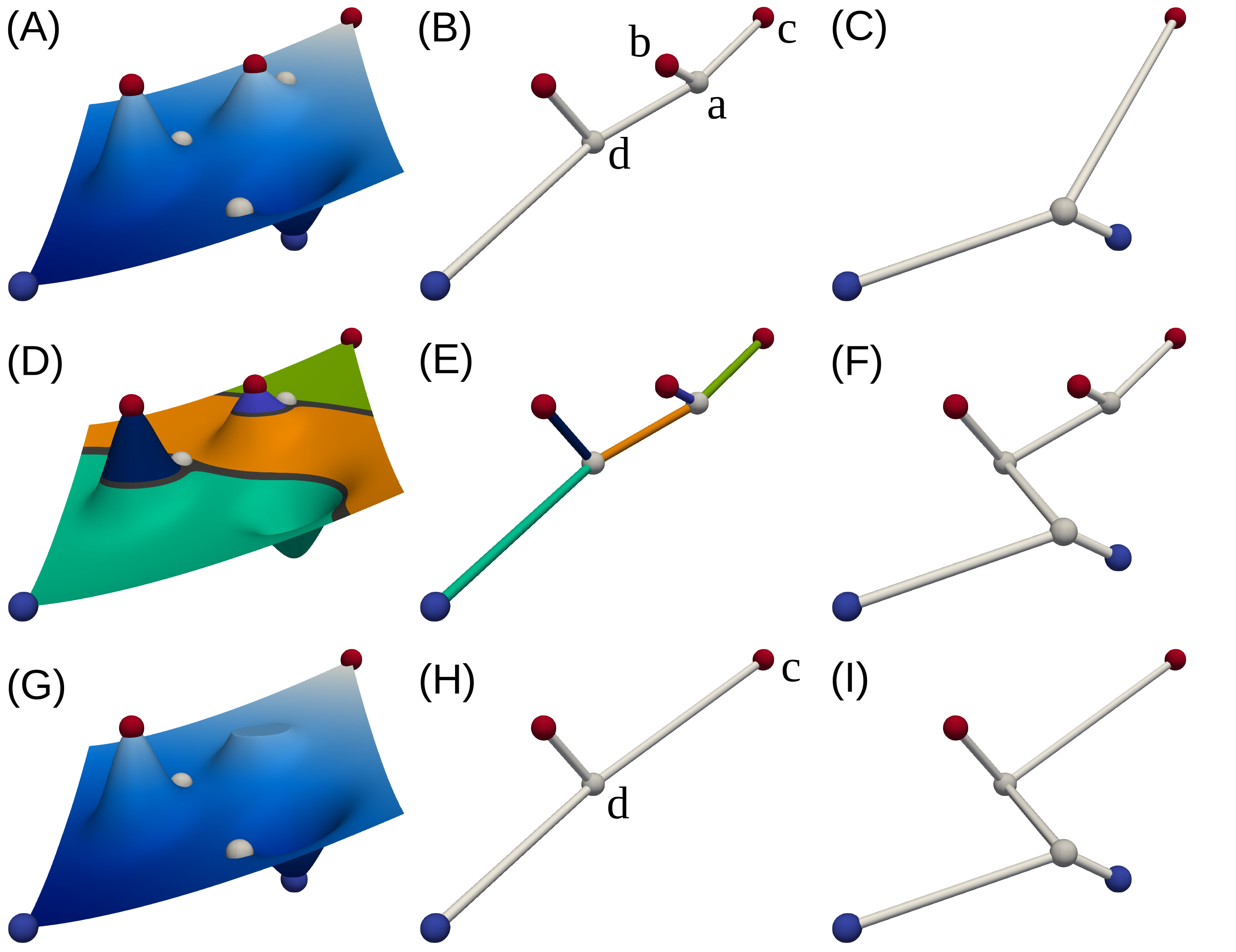}
    \vspace{-6mm}
    \caption{(A) Visualizing the graph of a 2D scalar field $f$ defined on a square domain. Local minima are in blue, saddles are in white, and local maxima are in red. (B) Split tree of f.  (C) Merge tree of $f$ (a.k.a.,~join tree). (D) The domain is colored according to split-tree-induced segmentation. (E) Split tree is colored according to the segmentation in (D). (F) Contour tree of $f$. (G) The graph of $f$ after its persistence simplification where a peak is removed. (H) Split tree of $f$ after persistence simplification. Edge $ab$ is removed. (F) Contour tree of $f$ after the merge tree in (B) has been simplified. }
    \label{fig:contour-tree}
    \vspace{-6mm}
\end{figure}

\para{Merge Tree.} 
Let $f:\X \rightarrow \R$ be a continuous scalar field defined on a simply connected domain $\X$. 
The \emph{sublevel set} of $f$ at a threshold $t \in \R$ is defined as 
$\X_t = f^{-1}(-\infty,t] := \{ x \in X \mid f(x) \leq t \}$. 
The \emph{merge tree} of $f$ tracks when (connected) components of $\X_t$ appear and merge as $t$ increases. 
$\X_t$ evolves from being an empty set to contain components surrounding various local minima; these components then merge into one other until eventually there is only a single component.
Leaves of the merge tree correspond to local minima, and interior nodes correspond to saddles where components merge. 
\Cref{fig:contour-tree}(A) and (C) visualize a scalar field and its merge tree.
Formally, we define an equivalence relation $\sim$ on $\X$. We say that $x \sim y$ if and only if $f(x) = f(y)=t$ and $x$ belongs to the same component of $\X_t$ as $y$. The merge tree of $f$ is defined by the quotient space $\X/{\sim}$. 

The merge tree of $f$ defined above is sometimes referred to as the \emph{join tree}, whereas the merge tree of $-f$ is called a \emph{split tree} (see~\cref{fig:contour-tree}(B) for its visualization). The merge tree naturally induces a segmentation of the domain. Let $\phi$ be the canonical map that maps each $x \in \X$ to its equivalence class $[x]$ under $\sim$. Then for each edge $e$ of the contour tree, $\phi^{-1}(e)$ is a monotonic region in $\X$. The inverse image of each edge partitions the domain, which is called the merge-tree-induced segmentation. See \cref{fig:contour-tree}(D) (cf.,~(E)) for a split-tree-induced segmentation.

\para{Contour Tree.} 
The \emph{level set} of $f$ at a threshold $t \in \R$ is $f^{-1}(t):= \{ x \in X \mid f(x) = t \}$.
Each component of $f^{-1}(t)$ is called a \emph{contour}. 
A \emph{contour tree} tracks the relations among contours as $t$ increases. 
Analogous to a merge tree, as $t$ increases, components of $f^{-1}(t)$ appear at local minima, disappear at local maxima, and join or split at saddles. 
Formally, we define another equivalence relation $\sim$ on $\X$, where $x \sim y$ if and only if $f(x) = f(y)$ and $x$ belongs to the same contour as $y$. 
The contour tree is the quotient space $T=T(\X,f):= \X / \sim$. 
There is a 1-1 correspondence between the local extrema of $\X$ and the leaves of $T$. The interior nodes of $T$ correspond to a subset of the saddle points of $\X$ \cite{reeb1946points}. \cref{fig:contour-tree}(F) visualizes the contour tree. Similar to the merge tree, we can define a contour-tree-induced segmentation.

The classic algorithm for computing the contour tree~\cite{carr2003computing} combines join and split trees together to form a contour tree, and comes with a state-of-the-art multi-core implementation~\cite{gueunet2017task}.
In this paper, we compute the contour tree of the input data to determine the initial point-wise error bound. We use the algorithm of Gueunet et al.~\cite{gueunet2017task} (with an in-house implementation) to compute join and split trees, and the algorithm of Carr et al.~\cite{carr2003computing} to combine them into the contour tree. 

The algorithm by Gueunet et al.~\cite{gueunet2017task} constructs the merge tree one edge at a time. First, starting from each minimum $m$, the algorithm visits points surrounding $m$ in an increasing order until a saddle $s$ is reached. The edge $ms$ is discovered and added to the merge tree; in other words, $m$ is \emph{grown} during this process.  
For each saddle $s$, once all edges that terminate at $s$ have been discovered, $s$ is grown until some new saddle $s'$ is reached, and the edge $ss'$ is discovered and added to the merge tree. This process repeats until the merge tree has been completely discovered. 
To grow from a minimum or a saddle, the points surrounding that minimum or saddle are stored in a heap to ensure that they are processed in an increasing order. The use of heaps presents the main performance bottleneck of the algorithm.

\para{Persistence simplification.} 
In practice, real-world data typically contain noise that creates many small branches in the merge or contour tree, where persistence simplification~\cite{edelsbrunner2001hierarchical} can be used to eliminate these small branches and thereby separate topological features from noise. 
In the context of merge trees, ordinary persistent homology pairs a local extremum (i.e.,~a peak or a valley in \cref{fig:contour-tree}(A)) with a nearby saddle and assigns the pair a value of \emph{persistence}, which describes the scale at which the pair disappears via a perturbation to the function. The persistence is equal to the absolute difference in function value between an extremum and its paired saddle. A function $f$ is simplified by perturbing the function values in order to cancel pairs of critical points below a certain persistence threshold $\varepsilon$. See \cref{fig:contour-tree}(G) for an example. The cancellation of one pair of critical points of $f$ corresponds to a branch being removed from the merge tree, giving rise to its simplification.
The simplified contour tree is more complex \cite{hristov2021w}, but it can be computed by combining the simplified join and split trees. 

In~\cref{fig:contour-tree}, (G) depicts the graph of a function after persistence simplification. For the split tree shown in (B), assuming a persistence threshold of $\varepsilon \geq |f(a)-f(b)|$, the edge $ab$ will be removed after a persistence simplification at $\varepsilon$. As a result, node $c$ is directly connected to node $d$ in the split tree (H), and the join tree in (C) remains unchanged. (I) depicts the contour tree produced by combining the simplified split tree in (H) with the join tree in (C).

\subsection{A Review on TopoSZ}
\label{sec:TopoSZ}

Our framework builds upon a few ingredients from TopoSZ~\cite{yan2023toposz}. 
TopoSZ, in turn, modifies the pipeline from the error-bounded lossy compressor SZ version 1.4 \cite{tao2017significantly}. 

Let $f$ represent the input scalar field, and $f'$ be the reconstructed scalar field (after compression and decompression). 
Let $T$ be the contour tree of $f$ and $T_\varepsilon$ the persistence simplified contour tree at a threshold of $\varepsilon$. Let $T'$ and $T'_\varepsilon$ be defined analogously for $f'$.

SZ1.4 allows the user to specify $\xi$, a pointwise error bound during compression. In turn, there are two user-defined parameters in TopoSZ: a persistence threshold $\varepsilon$, and a pointwise error bound $\xi$. Unlike TopoQZ, TopoSZ does not require that $\varepsilon < \xi$. 
TopoSZ guarantees the preservation of the persistence simplified contour tree during compression while maintaining the pointwise error bound. 
That is, it guarantees that $T_\varepsilon = T'_\varepsilon$, and $|f(x)-f'(x)| \leq \xi$ for each $x \in \X$.

\para{Linear-scaling quantization.} 
SZ version 1.4 introduces a linear-scaling quantization technique to ensure that a strict absolute error bound $\xi$ is maintained. This technique is implemented in TopoSZ.

For each point $x \in \X$ with a ground truth value $f(x)$, an initial guess for its value $g(x)$ (e.g., from a Lorenzo predictor; see below) is shifted by an integer multiple of $2\xi$ to obtain a new value $f'(x)$ such that $|f'(x) - f(x)| \leq \xi$. 

\begin{figure}[!ht]
  \vspace{-3mm}
  \centering
  \includegraphics[width=\columnwidth]{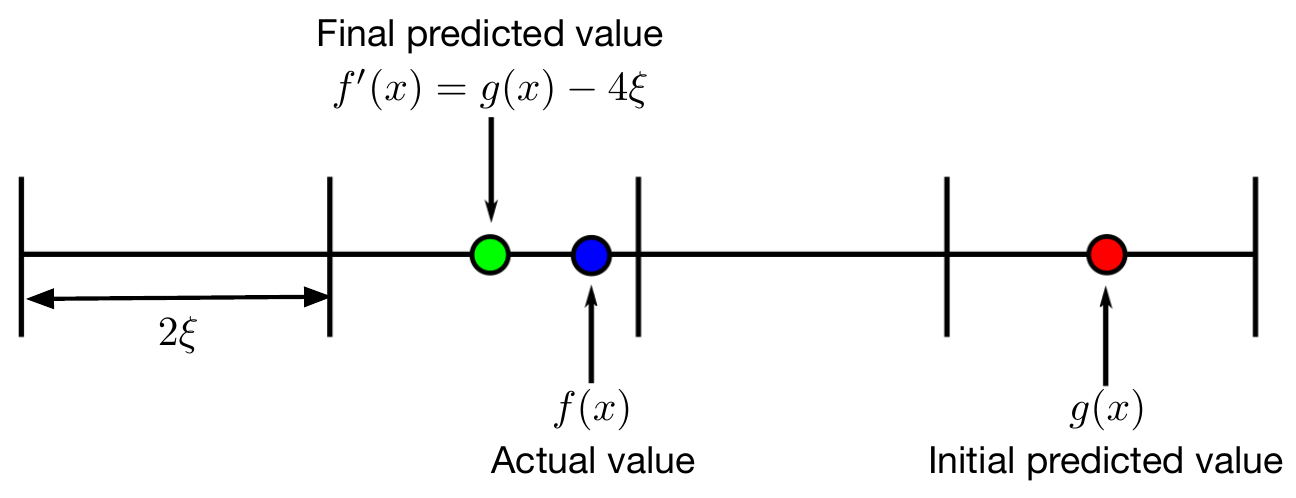}
  \vspace{-8mm}
  \caption{A standard implementation of a linear-scaling quantization.}
  \label{fig:linear-scaling-quantization}
  \vspace{-4mm}
\end{figure}

This process can be conceptualized as follows: divide the real line into intervals of length $2\xi$, where one interval is centered on $g(x)$. 
The compressor then calculates how many intervals to shift $g(x)$, so that it can assign a value to $f'(x)$ that is a distance less than $\xi$ from $f(x)$. By construction, if $f(x)$ lies in an interval of length $2\xi$ centered on $f'(x)$, then $|f(x)-f'(x)| \leq \xi$. 
This process is illustrated in \cref{fig:linear-scaling-quantization}.
\noindent Following this construction, each $x \in \X$ is assigned an integer $n_x$ (corresponding to how many intervals $g(x)$ is shifted) such that $f'(x) = g(x) + 2\xi n_x$. These quantization numbers $\{n_x\}$ are encoded and stored in the compressed file.

If the distribution of $\{n_x\}$ has low entropy (e.g.,~if $\{n_x\}$ are mostly zeros), then the quantization numbers can be compressed to small size using an entropy-based compression algorithm, such as Huffman coding. 
More accurate predictions $g(x)$ generally lead to $\{n_x\}$ with lower entropy.

\begin{figure}[!ht]
    \vspace{-2mm}
    \centering
    \includegraphics[width=\linewidth]{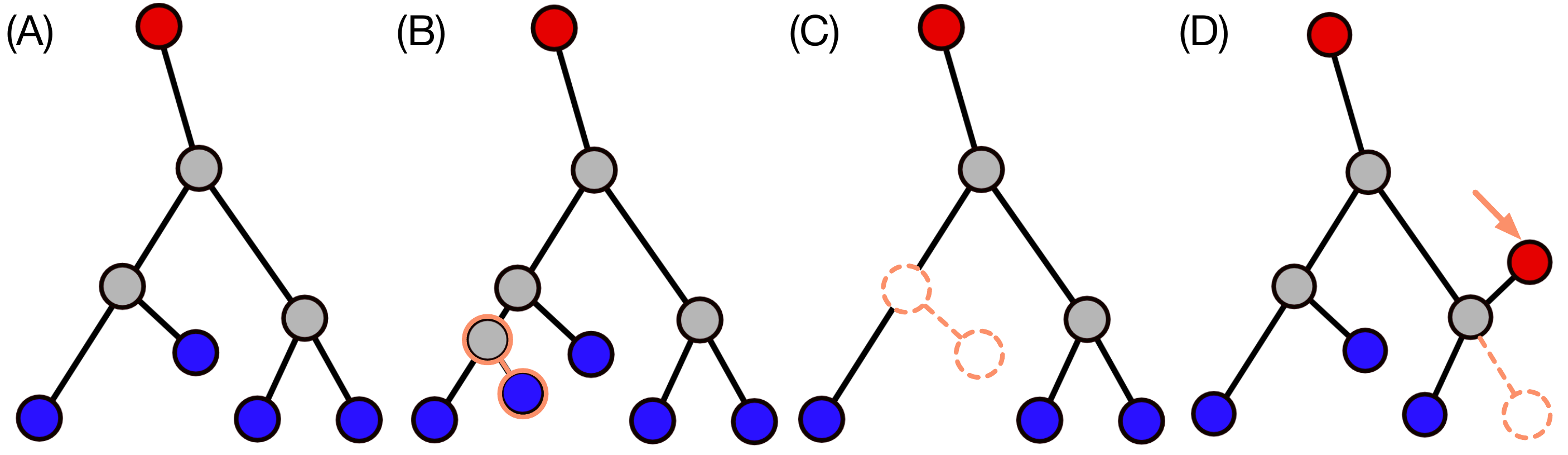}
    \vspace{-6mm}
    \caption{Three types of false cases. (A) The original contour tree. (B) A false positive: an extra edge is added. (C) A false negative: an edge is missing. (D) A false type: an edge contains a critical point as its endpoint that changes its type.}
    \label{fig:false-cases}
    \vspace{-4mm}
\end{figure}

\para{False Cases.} Yan et al.~\cite{yan2023toposz} introduced three types of false cases to quantify the level of contour tree preservation: false positives, false negatives, and false types, which are illustrated in \cref{fig:false-cases}.
A false positive occurs when a new edge appears in the contour tree of the reconstructed data that does not exist in the same position of the contour tree of the original data. A false negative occurs when an edge of the contour tree from the original data is missing from the contour tree of the reconstructed data. 
A false type occurs when the critical type (maximum, minimum, saddle) of one or both endpoints of an edge of the contour tree does not match between the original and reconstructed data.
TopoSZ focuses on false cases involving extremum-saddle pairs and its algorithm terminates when there are no such false cases. 

\para{TopoSZ Pipeline.} The TopoSZ pipeline is as follows:

\para{\underline{Step 1: Upper and lower bound calculation.}}
TopoSZ first applies persistence simplification to $f$ in order to calculate $T_\varepsilon$. For each point $x \in \X$, a lower bound $L(x)$ and an upper bound $U(x)$ are assigned to $x$ according to the contour-tree-induced segmentation.
If $x$ belongs to the segmented region corresponding to edge $ab \in T_\varepsilon$, then $L(x) = \min(f(a),f(b))$ and $U(x) = \max(f(a),f(b))$. The nodes of the contour tree are stored losslessly.

\para{\underline{Step 2: Prediction.}} 
TopoSZ uses a Lorenzo predictor \cite{ibarria2003out} to predict the values of each data point. For each point $x$, the Lorenzo predictor predicts $f(x)$ as a weighted sum of the values from previously predicted points $x'$ satisfying $\|x-x'\|_\infty = 1$. The weights are fixed, and are chosen such that quadratic functions will be perfectly predicted.

\para{\underline{Step 3: Linear-scaling quantization.}} TopoSZ uses linear-scaling quantization with a decreased interval size to ensure that the pointwise upper and lower bounds, as well as the global error bound $\xi$, are maintained for each $x \in \X$. For any $x \in \X$ where no possible quantization code $n_x$ satisfies these conditions, $f(x)$ is stored losslessly.

\para{\underline{Step 4: Iterative upper and lower bound tightening.}} 
If the results from Step 3 do not perfectly preserve the contour tree, that is, if there are false cases presented in the reconstructed data, then the upper and lower bounds are tightened around points corresponding to those false edges, and then Step 3 is repeated. This cycle repeats until there are no false cases. 
%The specifics of this step are reviewed in \cref{sec:topoSZ-detail}.

\para{\underline{Step 5: Lossless compression.}} The numbers from linear-scaling quantization are encoded using Huffman Coding. The relevant information is then stored in a binary file that is further compressed using ZSTD~\cite{collet2018zstandard}.
\section{Method}
\label{sec:method}
We give an overview of our framework in~\cref{sec:method-overview}.
We then describe two novel and technical ingredients in our framework: the logarithmic-scaling quantization (\cref{sec:augment-quantization}) and the progressive upper and lower bound tightening (\cref{sec:augment-tightening}).

\subsection{Overview}
\label{sec:method-overview}

We now describe our framework for augmenting any lossy compressor (called a \emph{base compressor}) to preserve contour trees and maintain strict error bounds. 
Our framework requires two user-specified parameters, a persistence threshold $\varepsilon$ and a pointwise absolute error bound $\xi$. 
It also requires user-specified parameters associated with the specific base compressor being augmented. Our implementation works with rectilinear meshes, and it could easily be modified to work with any simply-connected tetrahedral mesh.

Our framework guarantees that, for any augmented compressor, $T_\varepsilon = T_\varepsilon'$ and $|f(x)-f'(x)| \leq \xi$ for every $x \in \X$. Starting with a standard compressor as the base compressor, we start with a step-by-step overview of our framework. 

\para{\underline{Step 1: Upper and lower bound calculation.}}~We store critical points of the simplified contour tree $T_\varepsilon$ losslessly. We calculate the initial pointwise upper and lower bounds for other point $x \in \X$. The key idea is to locate an edge $ab$ in $T_{\varepsilon}$ whose corresponding range of function values contains $f(x)$. This requires a careful computation using the join and split trees of $T_{\varepsilon}$; see \cref{sec:algorithm-details} for details. 
We let $L(x) = \min(f(a),f(b)) + \zeta$ and $U(x) = \max(f(a),f(b))-\zeta$, where $\zeta = 10^{-5}|f(b)-f(a)|$.
If we allow $x$ to have the same function value as $a$ or $b$, the topology may be altered (e.g., along the boundary of the induced region), resulting in more false cases. Adjusting the error bound by $\zeta$ prevents such issues. We also adjust $L(x)$ and $U(x)$ as needed to ensure that if $L(x) \leq f'(x) \leq U(x)$ then $|f(x)-f'(x)| \leq \xi$. 
 
When computing $T_\varepsilon$, we compute the join and split trees of $f$ and simplify the trees directly with persistence threshold $\varepsilon$. We then combine them to obtain $T_\varepsilon$. During this construction, we track which edge of $T_\varepsilon$ each point $x \in X$ corresponds to. Compared to simplifying the entire scalar field $f$ and then computing the contour tree of the simplified field (like TopoSZ), our strategy leads to equivalent results in less time.

\para{\underline{Step 2: Base compressor.}} 
We apply the base compressor to the input data $f$. 
We compress and then decompress the data to assess changes that need to be made during decompression. 
We refer to the compressed-then-decompressed data as the \emph{intermediate data}.

\para{\underline{Step 3: Logarithmic-scaling quantization.}} 
We introduce a novel quantization technique that respects the pointwise upper and lower bounds imposed in Step 1. 
If possible, the entropy of the quantization numbers $\{n_x\}$ will be identical to that of standard linear-scaling quantization.
However, when linear-scaling quantization cannot produce a prediction for a point $x$ that respects $L(x)$ and $U(x)$, $x$ will be quantized with more precision (i.e.,~$\xi \leftarrow \xi/2$) to satisfy those bounds.

\para{\underline{Step 4: Progressive upper and lower bound tightening.}} 
We introduce a novel technique for calculating adjustments to the intermediate data to guarantee that the contour tree is preserved.
We compute the join and split trees directly. If a false edge is detected during computation, the upper and lower bounds are tightened around points in the segmentation region corresponding to the edge (see \cref{sec:merge-and-contour-tree}). All edges whose growth involved these points are recomputed.
We continue until the join and split trees of the decompressed data match those of the ground truth. We do not compute the contour tree directly as the preservation of the join and split trees guarantees the preservation of the contour tree.

\para{\underline{Step 5: Lossless compression.}} 
We encode the quantization numbers using Huffman coding. The output of the base compressor, the encoded quantization numbers, and any losslessly stored values are written to a binary file which is further losslessly compressed using xz, a general-purpose data compression tool available via {XZ Utils}~\cite{XZUtils}.

\subsection{Logarithmic-Scaling Quantization}
\label{sec:augment-quantization}

We now describe the first novel ingredient in our framework: a variable precision quantization technique that preserves tight pointwise upper and lower bounds. %without significantly compromising the entropy of the overall distribution of quantization numbers. 
For each $x \in \X$, the intermediate data contains an estimated value $g(x)$ for the ground truth value $f(x)$. 
Let $L(x)$ and $U(x)$ denote the lower and upper bounds assigned to $x$.
To ensure that $L(x) \leq f'(x) \leq U(x)$, we assign to each $x \in \X$ a numerator $a_x \in \Z$ and a precision $p_x \in \N$ that indicates the number of iterations. 
Our reconstructed value is 
\begin{equation}
f'(x) = g(x) + \frac{2\xi \cdot a_x}{2^{p_x}}.
\label{eq:fprime-original}
\end{equation}

To calculate each $a_x$ and $p_x$, we first set $p_x=0$. 
We then look for the value of $a_x$ satisfying 
\begin{equation*}
L(x) \leq g(x) + \frac{2\xi \cdot a_x}{2^{p_x}} \leq U(x)
\label{eq:Bounds}
\end{equation*}
such that $|a_x|$ is minimized. If there is no valid value of $a_x$, we increase $p_x$ by $1$ and search again. This process is repeated until a valid $a_x$ is found. If $p_x$ reaches an arbitrary threshold, we stop searching and instead store $f(x)$ losslessly. We set this threshold equal to $11$.

When $p_x = 0$, the above process is the same as the standard linear-scaling quantization, except that we also seek to maintain the upper and lower bounds. 
Each time a linear-scaling quantization fails to identify a valid choice for $a_x$ that yields a value of $f'(x)$ within the upper and lower bounds for $x$, we cut the interval lengths in half by increasing $p_x$ by $1$ and continue searching.
When the interval lengths are smaller, it is more likely that a valid choice of $a_x$ exists. 
It is also possible that during an iteration, multiple valid choices of $a_x$ exist, so we choose the one with the smallest absolute value to minimize the entropy of $\{a_x\}$. 

\begin{figure}[!ht]
    \centering
    \vspace{-2mm}
    \includegraphics[width=\linewidth]{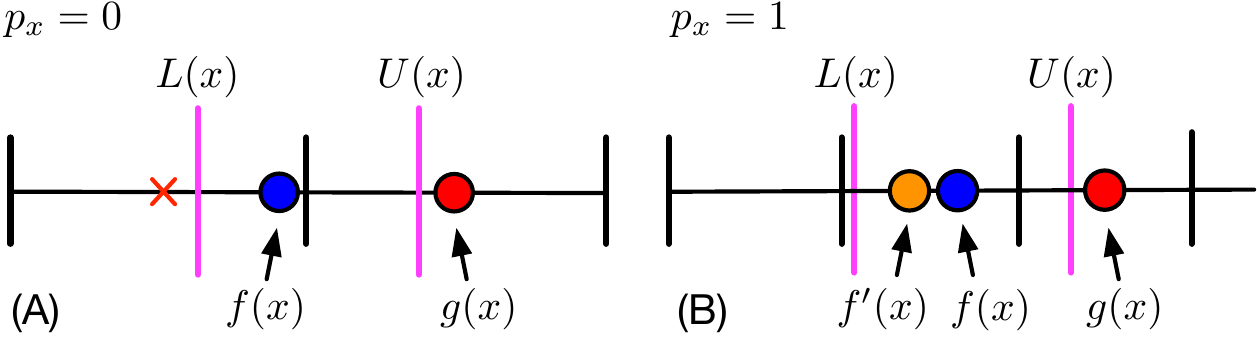}
    \vspace{-6mm}
    \caption{(A) If $p_x = 0$, there are no valid quantization intervals. (B) Increasing $p_x$ to $1$ allows for a valid quantization interval.}
    \label{fig:log-scale-quantization}
    \vspace{-2mm}
\end{figure}

This process is illustrated in \cref{fig:log-scale-quantization}. 
(A) contains an example where there are no quantization intervals where we can place $f'(x)$ to respect the upper and lower bounds. 
In (B), by raising the precision $p_x$ by 1, the quantization intervals are halved, giving a valid choice for $f'(x)$.

When encoding the data, we store a single quantization number $n_x$ for each $x \in \X$. 
To calculate each $n_x$, we first find the maximum precision $p_m$ used for any single point. The points are assigned the single quantization number $n_x = a_x \cdot 2^{p_m-p_x}$ and the max precision $p_m$ is stored in the compressed output. 
During decompression, the point $x$ is assigned the value 

\begin{equation}
f'(x) = g(x) + \frac{2\xi \cdot n_x}{2^{p_m}}.
\label{eq:fprime}
\end{equation}
Setting $n_x = a_x \cdot 2^{p_m-p_x}$ in Eq.~\eqref{eq:fprime} means that
\begin{equation*}
  g(x) + \frac{2\xi \cdot n_x}{2^{p_m}} = g(x) + \frac{2\xi \cdot a_x \cdot 2^{p_m-p_x}}{2^{p_m}} = g(x) + \frac{2\xi \cdot a_x}{2^{p_x}}.
  \label{eq:logscale}  
\end{equation*}
Therefore, the formulation in Eq.~\eqref{eq:fprime} is equivalent to the original formulation of $f'$ in Eq.~\eqref{eq:fprime-original}.

In comparison with TopoSZ, the above variable precision technique allows us to store fewer points losslessly.
In order to ensure the quantization numbers do not get too large, if any point has a precision greater than $10$ it is stored losslessly. This ensures that $p_m \leq 10$ for all trials.

\subsection{Progressive Upper and Lower Bound Tightening}
\label{sec:augment-tightening}

We now describe the second novel ingredient in our framework, namely, a \emph{progressive error bound tightening} process. 
Specifically, the process computes the join and split trees of the decompressed data. During the computation, it detects false cases, and tightens the upper and lower bounds in the neighborhoods of false cases. The algorithm progresses through merge tree computation, checking the correctness of each edge and tightening when needed, until every edge is correctly preserved.
The process allows us to bypass iteratively recomputing the entire contour tree (in the case of TopoSZ), significantly speeding up the compression process. During the tightening process, we work with merge trees (instead of contour trees), since the persistence of a leaf (local extremum) can be computed from its nearby saddle based on branch decomposition (i.e.,~local information), thereby allowing for our progressive tightening strategy. By contract, computing the persistence of a leaf of a contour tree may require global information from the whole contour tree due to the existence of V and W structures~\cite{hristov2021w}.

We describe this process for the join tree, which works analogously for the split tree. We only consider false cases involving extremum-saddle pairs. 

\para{False case detection}. To detect false cases, we construct $T'$. Doing so allows us to locate mismatches between edges in $T'_\varepsilon$ and those in $T_\varepsilon$.
We construct $T'$ using a modified version of the edge growing procedure from local minima and saddles (see~\cref{sec:merge-and-contour-tree}).
To start, we extract a list of local minima of $f'$ sorted by decreasing function values. Then, proceeding in sorted order, we grow an edge from each local minimum $m$ to a saddle $s$, and check two cases for $s$; see \cref{sec:algorithm-details} for illustrations: 

\underline{Case (I).} If $s$ is unpaired, i.e., $m$ is the first local minimum (among all local minima) whose growth terminates at $s$, then $m$ and $s$ form a persistence pair, with a persistence $p =|f'(s)-f'(m)|$. 
If $p < \varepsilon$, then the edge $ms$ does not belong to $T'_\varepsilon$; otherwise, $ms$ belongs to $T'_\varepsilon$.  

\underline{Case (II).} If $s$ is already paired, then $m$ must pair with some other saddle $s'$, and $s'$ must be an ancestor of $s$ in the join tree. A paired $s$ means that $s$ has been discovered earlier during the growth of another local minimum $m'$ such that $m'$ and $s$ form a persistence pair with persistence $p'$, and the edge $m's$ belongs to $T'$. 

\underline{Case (II.a).} 
Suppose that $p' \geq \varepsilon$. Since $m'$ preceds $m$ in the sorted order, $f'(m') > f'(m)$. Since $s'$ is an ancestor of $s$, $f'(s') > f'(s)$. Therefore $|f'(s') - f'(m)| > |f'(s) - f'(m')| = p' \geq \varepsilon$. 
Thus, the pair $(m,s')$ has a persistence above $\varepsilon$, and $ms$ must be an edge in $T'_\varepsilon$.

\underline{Case (II.b).} 
Now suppose that $p' < \varepsilon$. In this case, we do not have enough information to determine the persistence of $(m,s')$. Therefore, we grow from saddle $s$ to reach a new saddle $s''$. We then check cases (I) and (II) again, using $s''$ in place of $s$. 

Once we are done checking cases (I) and (II), if $m \notin T'_{\varepsilon}$ but $m \in T_{\varepsilon}$, then $m$ is a false negative. 
Likewise, if $ms \in T'_{\varepsilon}$ but $ms \notin T_{\varepsilon}$, then $ms$ is a false positive. 

Growing the global minimum will never produce a false case as long as the rest of $T'_\varepsilon$ is correctly predicted. Thus, we skip the growth at the global minimum, denoted as $\hat{m}$. 
Because $\hat{m}$ is the last growth that remains active, its growth will form the \textit{trunk}, a monotone sequence of edges to the root that links $\hat{m}$ to the remaining saddles~\cite{gueunet2017task}. Since $\hat{m}$ and the remaining saddles are already correctly predicted, so is the trunk, therefore no further false cases are possible, and we skip growing $\hat{m}$. 
This algorithm also admits a number of special cases; see~\cref{sec:algorithm-details}.

\para{Progressive false case correction.} 
If there is a false case, we first tighten the upper and lower bounds of points in some region $R$ to correct it. If $ms$ is a false positive, then $R$ is the region of the merge-tree-induced segmentation of $f'$ corresponding to $ms$. If $m$ is a false negative, and edge $m\hat{s}$ belongs to $T_\varepsilon$ (for some saddle $\hat{s}$), then $R$ is the region of the merge-tree-induced segmentation of $f$ corresponding to $m\hat{s}$. If the same false case occurs multiple times, we grow the region $R$. We tighten the upper and lower bounds of each $x \in R$ similarly to TopoSZ, but we tighten more aggressively to speed up compression. 
We then update the decompressed data $f'$ to respect the new bounds; see~\cref{sec:algorithm-details} for numerical specifics and a comparison with TopoSZ.

Once we update $f'$, these updates may affect parts of the join and split trees beyond the false cases, thus we must recompute those areas to ensure correctness. Specifically, we must check for any extrema bordering $R$ that may have appeared or disappeared as a result of the tightening process and update the trees accordingly. Let $E$ be the set of edges whose segmentation regions border $R$. Then the tightening also may have affected each edge $e \in E$ and every ancestor of $e$ (i.e.,~edges
connecting $e$ to the root of the tree). We recompute all such edges to ensure correctness. As before, we recompute parts of the tree in order of the function values.

\section{Experimental Results}
\label{sec:results}

\begin{figure*}[!ht]
    \centering
    \begin{subfigure}{0.02\textwidth}
    \raisebox{0.1\height}{\includegraphics[angle=90,width=\textwidth]{fig-colorBar}}
    \end{subfigure}
    \begin{subfigure}{0.92\textwidth}
    \includegraphics[width=\textwidth]{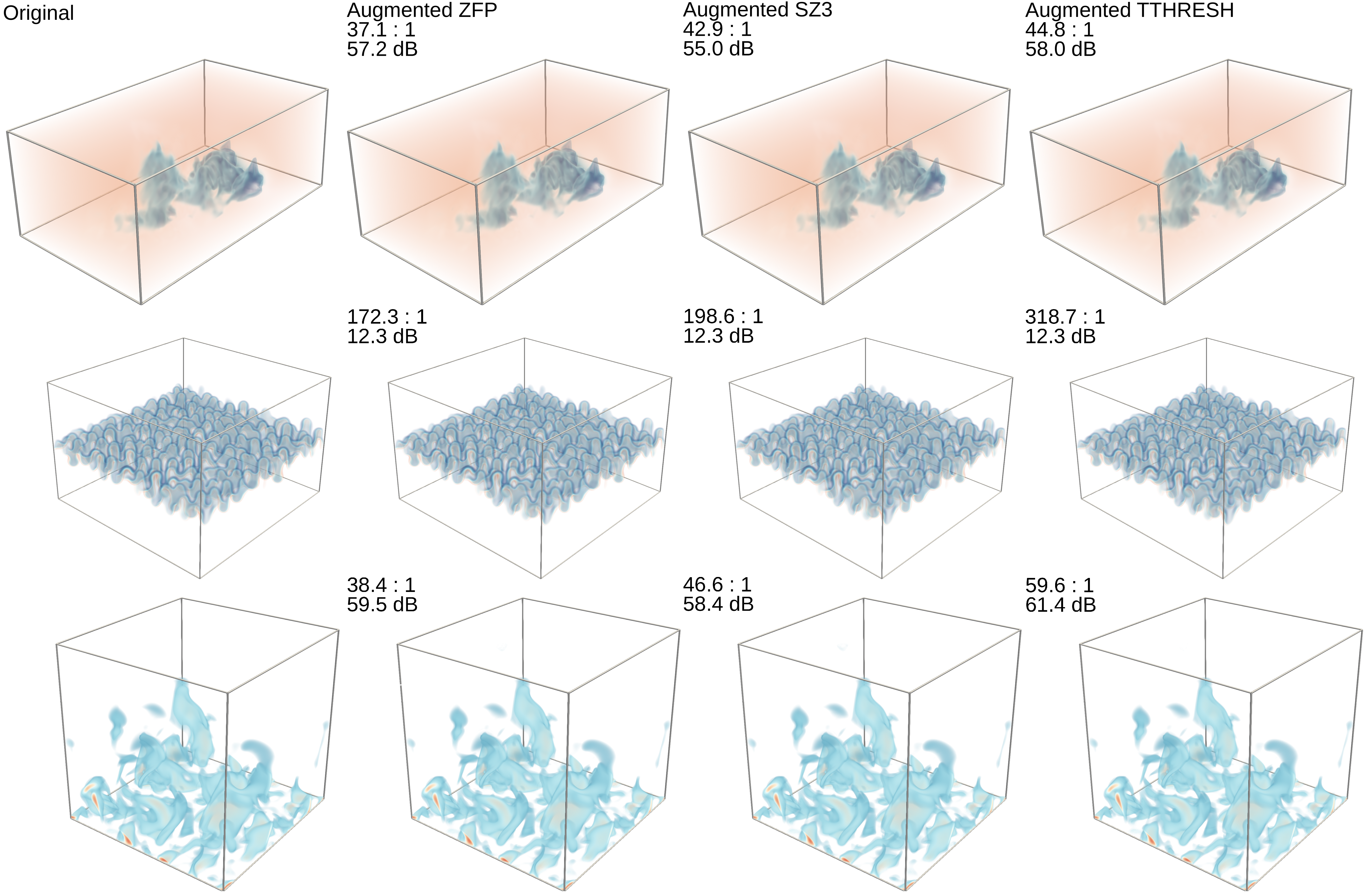}
    \end{subfigure}
    \vspace{-3mm}
    \caption{Scientific datasets compressed using different augmented compressors with topological controls. From left to right: the original input dataset, the reconstructed datasets from Augmented ZFP, Augmented SZ3, and Augmented TTHRESH, 
    respectively, that preserve the contour trees up to a persistence threshold $\varepsilon = 0.04$. From top to bottom: Tangaroa, Miranda, S3D datasets, respectively. We also display the PSNR and compression ratio next to each decompressed dataset.}    
    \label{fig:volume-render}
    \vspace{-6mm}
\end{figure*}

We provide an overview in~\cref{sec:results-overview}, describing the base compressors and datasets used in our experiments, highlighting the main takeaways, and introducing the evaluation metrics. 
We include compressor configurations and implementation details in~\cref{sec:configurations}. 
In \cref{sec:augmented-compressors} we describe the main utilities of our augmented compressors in preserving contour trees in the reconstructed data.
We evaluate a number of augmented compressors qualitatively and quantitatively, followed by a comparison against the state-of-the-art topology-preserving compressors in \cref{sec:compare-topology}.
We end this section with a run time analysis in \cref{sec:run-time}.

%---------------------------------------
\subsection{An Overview of Experiments}
\label{sec:results-overview}

We present a comparative analysis of five error-bounded lossy compressors augmented with our framework, including the classic compressors ZFP \cite{lindstrom2014fixed}, SZ3 \cite{liang2022sz3}, and TTHRESH \cite{ballester2019tthresh}, a custom-built cubic spline interpolation (CSI) model, and a deep learning-based compressor Neurcomp \cite{lu2021compressive}. 
We test these augmented compressors---denoted as Augmented ZFP, Augmented SZ3, and so on---against two state-of-the-art topology-preserving compressors, TopoSZ~\cite{soler2018topologically} and TopoQZ~\cite{yan2023toposz}. 

We test the five augmented compressors and two topology-preserving compressors on nine volumetric datasets from scientific simulations. The Nyx dataset is very topologically complex---its contour tree has over twenty million nodes---and it is included as a stress test. See \cref{tab:datasets} and \cref{sec:datasets} for details on these datasets.

We further conduct an ablation study demonstrating the effectiveness of logarithmic-scaling quantization and progressive error bound tightening. In every trial, logarithmic-scaling quantization leads to higher compression ratios, whereas progressive tightening results in faster compression times. We also analyze the individual effects of varying $\varepsilon$ and $\xi$; see~\cref{sec:other-experiments} for details on these experiments and the ablation study.

\begin{table}[!ht]
\scriptsize
\centering{
\begin{tabu}{c|*{2}{c}}
\toprule
\textbf{Dataset}  & \textbf{Dimension} & \textbf{Size (MB)}  \\ 
\midrule
QMCPACK      & $69 \times 69 \times 115$          & 4.4         \\
Tangaroa     & $300 \times 180 \times 200$         & 27.0        \\
Earthquake   & $175 \times 188 \times 50$          & 28.2        \\
Ionization   & $310 \times 128 \times 128$         & 40.6       \\
Isabel       & $500 \times 500 \times 90$          & 105.0     \\
Miranda      & $384 \times 384 \times 256$         & 302.0      \\
Nyx          & $512 \times 512 \times 512$         & 641.4      \\
S3D          & $500 \times 500 \times 500$         & 1000.0    \\
SCALE-LETKF  & $1200 \times 1200 \times 98$        & 1129.0    \\
\bottomrule
\end{tabu}
}
\vspace{-2mm}
\caption{Scientific datasets used for compression analysis.}
\label{tab:datasets}
\vspace{-4mm}
\end{table}

\para{Highlighted results.}
We highlight our experimental results below. 
\begin{itemize}[noitemsep,leftmargin=*]
\item Applying any of the five original base compressors to any of the nine datasets produces a large number of topological false cases in the reconstruction, even with a small pointwise error bound. On the other hand, augmenting any compressor with our general framework completely eliminates these false cases while maintaining a user-specified error bound (\cref{sec:augmented-compressors}).
\item Augmented TTHRESH and Augmented ZFP respectively yield the best compression ratios and run times among all the augmented compressors (\cref{sec:augmented-compressors}).
\item Our augmented compressors generally have a better trade off between bit-rate and reconstruction quality compared to TopoQZ and TopoSZ while taking similar or less time to run (\cref{sec:compare-topology}).
\item Our framework has a worst-case time complexity of $O(F h n \log n)$, where $h$ is the maximum height of the contour tree during tightening and $F$ is the number of false cases during computation. The majority of the compression time is spent on computing merge trees (\cref{sec:run-time}).
\end{itemize}

\para{Evaluation metrics.} 
We evaluate whether the contour tree has been perfectly preserved in the reconstructed (decompressed) data. 
We also evaluate the standard compression metrics of compression ratio, bit-rate, and peak signal-to-noise ratio (PSNR).
We further employ topology-based metrics of the bottleneck distances $d_B$~\cite{cohen2005stability} and the Wasserstein distances $d_W$ \cite[page 183]{edelsbrunner2022computational} to quantify the topological similarity between the original data and the reconstructed data. 
The evaluation metrics are described in detail in \cref{sec:evaluationMetrics}.

In general, higher values of PSNR indicate better reconstruction quality, and lower values of $d_B$ and $d_W$ indicate higher topological similarity. 
We measure the total compression time for each compressor, which includes (a) the total time to run the base compressor, and (b) the time to augment the output of the base compressor. We also measure decompression time for each compressor. We measure compression and decompression time for TopoSZ and TopoQZ as well. For our framework and TopoSZ, we decompress to RAW binary format. Because TopoQZ is integrated in the Topology Toolkit, an extension for ParaView, we decompress to VTK image format.

% ---------------------------------------
\subsection{Compressor Configurations and Implementation}
\label{sec:configurations}

In addition to augmenting the out-of-box base compressors SZ3, TTHRESH, ZFP, and Neurcomp, we implement and augment our own super-resolution compressor, a simple custom-built cubic spline interpolation (CSI) model.
It compresses a dataset by downsampling the data by a user-defined ratio in each direction (called a target scale factor) and uses a cubic spline interpolation technique for reconstruction that is similar to the one implemented in SZ3.
%We also considered the Sliced Wasserstein Autoencoder \cite{kolouri2018sliced} used in the AE-SZ compressor \cite{liu2021exploring}. 
%However, this model is excluded from our experiments since it performed significantly worse than the other compressors during initial tests.   

We compare our augmented compressors to TopoSZ and TopoQZ. We use the TopoQZ implementation in TTK~\cite{TiernyFavelierLevine2017}. 

Our general framework requires two user-defined parameters, a persistence threshold $\varepsilon$ and a global absolute pointwise error bound $\xi$.
$\varepsilon$ represents, as a percentage of the range, the level of persistence  simplification. 
For example, $\varepsilon = 0.01$ corresponds to a persistence simplification by $1\%$ of the range of the scalar function. Similarly, $\xi$ is the percentage of the range that will be used as an absolute error bound.

Each base compressor takes a number of intrinsic parameters in order to run. 
Both ZFP and SZ3 require an absolute error bound, denoted as $\delta$ and $\eta$, respectively. 
CSI requires a target scale factor $s$.
TTHRESH takes in a target RMSE of $\tau$. 
Neurcomp requires a target compression ratio $c$. 
Changing the intrinsic parameters of a base compressor will cause it to generate different intermediate data which will be augmented differently. 
As a result, even though our augmented compressor guarantees topology preservation and maintains the user-defined global error bound, the compression results may vary. 

For our experiments, we set the error parameter for each base compressor (except CSI and Neurcomp) to be equal to $k\xi$ for some $k \in \R$ that is compressor-dependent. Specifically, we set $\delta = 5\xi$, $\eta = 0.25\xi$, and $\tau = 0.05\xi$. To decide each value of $k$, we conduct a grid search and observe the effects of different values of $k$ across different values of $\xi$ and different datasets. The optimal value of $k$ varies between datasets and values of $\xi$, but the values that we chose are always approximately optimal. Hypothesized explanations as to why these values of $k$ improve results are described in \cref{sec:base-compressor-parameters}. We also set $c = 100$ and $s = 7$. We chose these configurations because they empirically led to the highest compression ratios.

To differentiate from the persistence threshold $\varepsilon$ used by an augmented compressor, TopoQZ takes a persistence threshold $e$. The TTK implementation of TopoQZ is tightly coupled with ZFP, which requires an error bound $\zeta$, allowing for a total pointwise error upper-bounded by $e+\zeta$.
To measure compression ratio and compression times while respecting a topological constraint $\varepsilon$ and an error bound $\xi$, we measure how each augmented compressor and TopoSZ perform for $\varepsilon = 0.04$ and $\xi = 0.012$. When testing TopoQZ, to ensure that it respects both $\varepsilon$ and $\xi$, we set $e = \zeta = 0.006$ so that the max error is less than $0.012$.

To measure the trade off between compression ratio and reconstruction quality, for TopoSZ and each augmented compressor, we set $\varepsilon = 0.04$ and vary $\xi \in \{$0.003, 0.006, 0.009, 0.012, 0.015, 0.018$\}$. In some cases we need to vary $\xi$ (and $\zeta$, for TopoQZ) in a different range in order to obtain a meaningful curve. Notably, for TopoQZ, we set $e = 0.04$ and vary $\zeta \in \{$0.003, 0.11, 0.22, 0.33, 0.44, 0.55$\}$. We further discuss our methodology for choosing parameters and provide the parameter used in \cref{sec:reconstruction-quality-extra}. 

The combination of a chosen compressor, a fixed dataset, a value of $\varepsilon$ and $\xi$, is a trial.
We perform each trial on a single cluster node running an Intel Xeon Sandy Bridge-E processor with 16 cores and 64GB of RAM.
For Neurcomp, we perform the training on an RTX 2080ti GPU with 32GB of RAM.

\begin{figure}[!ht]
\begin{subfigure}{0.02\linewidth}
\raisebox{2.2\height}{
\includegraphics[angle=90, width=\linewidth]{fig-colorBar.png}}
\end{subfigure}
\begin{subfigure}{0.97\linewidth}
    \includegraphics[width=\linewidth]{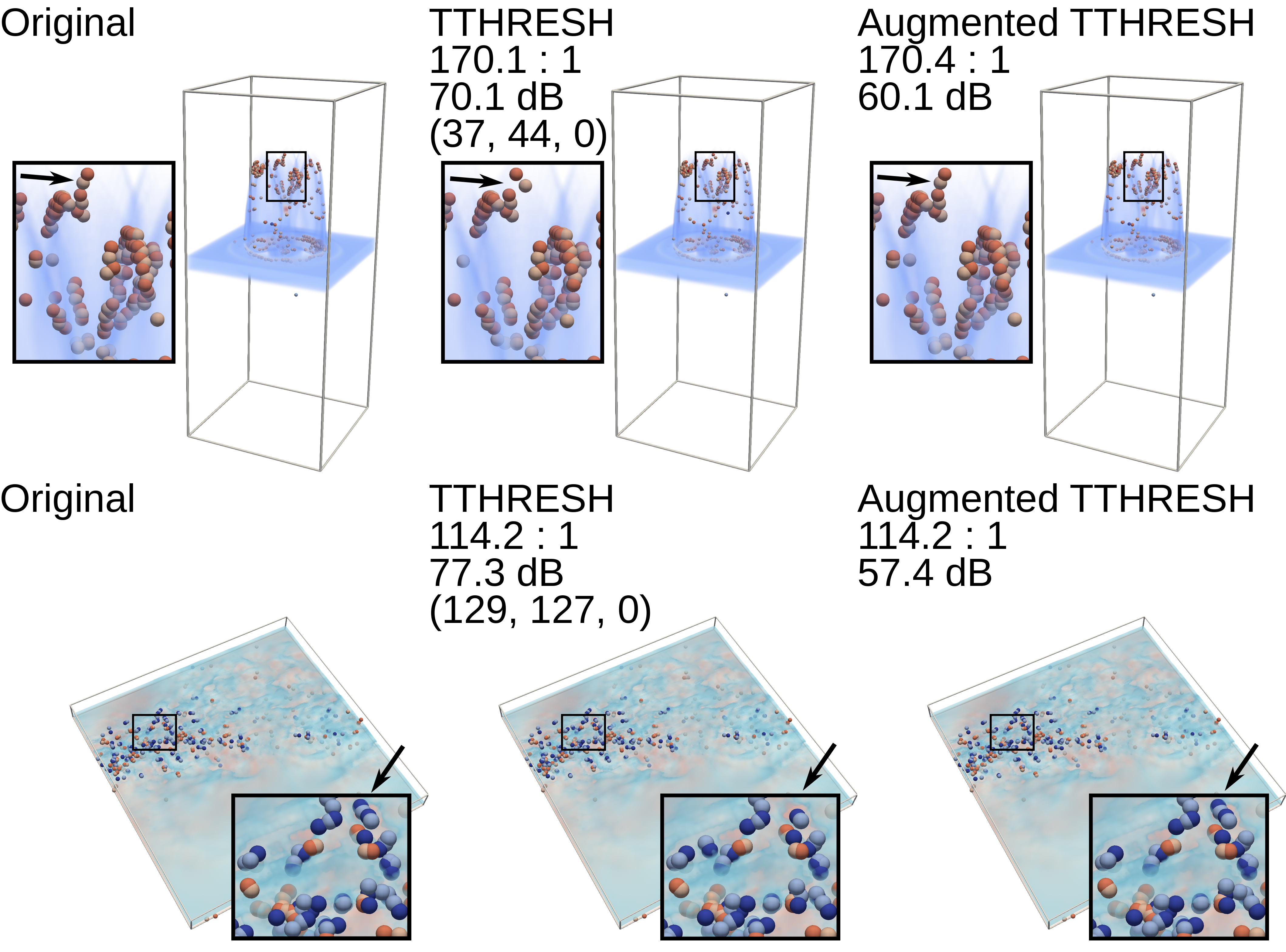}
\end{subfigure}
\vspace{-6mm}
\caption{Zoomed-in views of the critical points of the contour trees of the Ionization (top) and SCALE-LETKF (bottom) datasets with persistence simplification $\varepsilon = 0.04$. For each dataset, arrows indicate one example where compression with TTHRESH led to critical points shifting. Local maxima are in orange, local minima are in dark blue, 1-saddles are in light blue, 2-saddles are in light orange. Top row: Ionization. Bottom row: SCALE-LETKF.}
\label{fig:zoom}

\centering{
\begin{tabular}{c|cc|c}
\hline
Dataset    & ZFP       & TTHRESH       & Total \#edges \\ \hline
QMCPACK    & (23,23,0) & (8,8,0)     & 69          \\ 
Tangaroa   &  (90, 92, 0) & (39, 46, 0) & 418         \\ 
Earthquake & (20, 19, 0) & (26, 25, 0)     & 169         \\
Ionization & (181, 187, 0)  & (37, 44, 0) & 568         \\
Isabel     & (16, 15, 0) & (15, 16, 0)  & 29          \\
Miranda    & (6, 6, 0) & (3, 3, 0)     & 11          \\ 
Nyx        & (133, 132, 0) & (2, 373, 0) & 743 \\
S3D       & (100, 100, 0) & (74, 80, 0) & 1013 \\
SCALE-LETKF & (152, 153, 0) & (129, 127, 0) & 401 \\ \hline
\end{tabular}
}
\vspace{-2mm}
\captionof{table}{Reporting the number of false cases (false positives, false negatives, false types) produced by base compressors SZ3 and TTHRESH, respectively, together with the total number of edges of the input (ground truth) contour tree. Contour trees are simplified with $\varepsilon=0.04$.}
\label{tab:base-false-cases}
\vspace{-6mm}
\end{figure}

%---------------------------------------
\subsection{Comparative Analysis of Augmented Compressors}
\label{sec:augmented-compressors}

In this section, we perform a comparative analysis of five augmented compressors, qualitatively and quantitatively. 
We visualize three scientific datasets before and after compression with {three of our augmented compressors} in \cref{fig:volume-render}. We also display the PSNR and compression ratio next to each decompressed dataset. 
Compression ratios and times for a single combination of $\varepsilon$ and $\xi$ are reported in \cref{tab:compression-task}. Charts showing the reconstruction quality on two datasets is reported in \cref{fig:reconstruction-quality}. Similar charts for the remaining datasets and compressors are shown in \cref{sec:reconstruction-quality-extra}. Results demonstrating the effect of independently varying $\varepsilon$ or $\xi$ on the evaluation metrics are given in \cref{sec:other-experiments}.

\subsubsection{Topological Guarantees}
When compressing a dataset with any base compressor, the contour tree of the data is often significantly distorted with a large number of false cases, whereas it is always perfectly preserved using our augmented compressor. This observation has been validated empirically in every trial: the contour tree is perfectly preserved in terms of the locations of its critical points and their connectivity.

For instance, we visualize the Isabel dataset in \cref{fig:teaser} using TTHRESH and augmented TTHRESH. We highlight parts of the contour trees via zoomed-in views before and after compression. In a bottom zoomed-in view, TTHRESH (middle) fails to preserve a few critical points of the contour tree. In a top zoomed-in view, TTHRESH (middle) preserves the locations of the critical points, but not their connectivity. In contrast, the augmented TTHRESH preserves both the locations and connectivity among the critical points.

We further provide zoomed-in views for the Ionization and SCALE-LETKF datasets in~\cref{fig:zoom}. We observe clearly that TTHRESH fails to predict many critical points, whereas augmented TTHRESH preserves them all.

In \cref{tab:base-false-cases}, we report the number of false cases, including both extremum-saddle and saddle-saddle connections in the contour tree reconstructed with ZFP and TTHRESH. We again use parameter configurations that produce the same compression ratios as their augmented versions with $\varepsilon = 0.04$ and $\xi = 0.012$ (for those configurations see \cref{sec:base-compressor-parameters}). In \cref{tab:base-false-cases}, we can see that ZFP and TTHRESH produce many false cases.

\begin{table}[!t]
\setlength{\tabcolsep}{2pt}
\resizebox{\columnwidth}{!}{
\begin{tabular}{cccccccc}
\hline
Dataset     & A-ZFP            & A-SZ3   & A-CSI         & A-TTHRESH      & \multicolumn{1}{c|}{A-Neurcomp} & TopoQZ             & TopoSZ          \\ \hline
\multicolumn{8}{c}{Compression Ratio}                                                                                                                \\ \hline
QMCPACK     & 58.7             & 86.1    & 102.3           & \textbf{104.8} & \multicolumn{1}{c|}{23.9}       & 23.4               & 27.8            \\
Tangaroa    & 37.3             & 43      & 33.5            & \textbf{44.8}  & \multicolumn{1}{c|}{15.3}       & --                 & 24.3            \\
Earthquake  & 86.1             & 127.4   & 79.4            & \textbf{129.2} & \multicolumn{1}{c|}{63.5}       & 13.4               & 50.1            \\
Ionization  & 118.8            & 121.5   & 119.9           & \textbf{170.5} & \multicolumn{1}{c|}{72.7}       & 30.0               & 25.1            \\
Isabel      & 47.4             & 103.5   & 70.6            & \textbf{182.2} & \multicolumn{1}{c|}{41.6}       & --                 & 37.6            \\
Miranda     & 172.3            & 198.6   & 157.2           & \textbf{318.7} & \multicolumn{1}{c|}{95.0}       & 76.5               & 95.9            \\
Nyx         & 65.3             & 69.5    & 70.4            & \textbf{84.5}  & \multicolumn{1}{c|}{18.9}       & --                 & --              \\
S3D         & 38.4             & 46.6    & 43.6            & \textbf{59.6}  & \multicolumn{1}{c|}{6.0}        & 9.2                & --              \\
SCALE-LETKF & 69.5             & 74.4    & 58.5            & \textbf{114.2} & \multicolumn{1}{c|}{8.6}        & 11.4               & --              \\ \hline
\multicolumn{6}{c|}{Total Compression and Augmentation Time}                                                  & \multicolumn{2}{c}{Compression Time} \\ \hline
QMCPACK     & 1.27             & 1.36    & 1.21            & 1.45           & \multicolumn{1}{c|}{172.08}     & \textbf{1.05}      & 10.46           \\
Tangaroa    & 9.51             & 10.99   & \textbf{9.33}   & 11.98          & \multicolumn{1}{c|}{1519.21}    & --                 & 314.56          \\
Earthquake  & \textbf{7.08}    & 7.39    & \textbf{7.08}   & 8.66           & \multicolumn{1}{c|}{1039.94}    & 8.08               & 48.17           \\
Ionization  & \textbf{8.68}    & 10.00   & 14.08           & 15.16          & \multicolumn{1}{c|}{1221.12}    & 10.40              & 425.31          \\
Isabel      & \textbf{33.77}   & 35.49   & 42.69           & 42.67          & \multicolumn{1}{c|}{7147.86}    & --                 & 367.10          \\
Miranda     & 223.52           & 284.18  & 248.51          & 348.72         & \multicolumn{1}{c|}{9359.59}    & \textbf{160.60}    & 434.98          \\
Nyx         & \textbf{1059.06} & 1137.33 & 5664.46         & 25594.54       & \multicolumn{1}{c|}{38959.73}   & --                 & --              \\
S3D         & 209.83           & 253.82  & \textbf{173.09} & 253.72         & \multicolumn{1}{c|}{34610.78}   & 633.13             & --              \\
SCALE-LETKF & \textbf{221.49}  & 399.32  & 343.58          & 371.19         & \multicolumn{1}{c|}{40887.62}   & 524.05             & --              \\ \hline
\multicolumn{8}{c}{Decompression Time}                                                                                                               \\ \hline
QMCPACK     & 0.14             & 0.32    & 0.14            & 0.17           & \multicolumn{1}{c|}{4.24}       & 0.63               & \textbf{0.01}   \\
Tangaroa    & 0.49             & 0.52    & 0.51            & 0.96           & \multicolumn{1}{c|}{16.32}      & --                 & \textbf{0.12}   \\
Earthquake  & 0.38             & 0.41    & 0.37            & 0.55           & \multicolumn{1}{c|}{8.50}       & 5.50               & \textbf{0.07}   \\
Ionization  & 0.48             & 0.54    & 0.54            & 0.83           & \multicolumn{1}{c|}{11.62}      & 5.08               & \textbf{0.10}   \\
Isabel      & 1.43             & 1.35    & 1.34            & 2.64           & \multicolumn{1}{c|}{79.61}      & --                 & \textbf{0.41}   \\
Miranda     & 2.92             & 3.1     & 3.38            & 4.53           & \multicolumn{1}{c|}{175.61}     & 53.2               & \textbf{0.67}   \\
Nyx         & \textbf{6.72}    & 8.05    & 9.71            & 9.49           & \multicolumn{1}{c|}{2457.91}    & --                 & --              \\
S3D         & \textbf{11.52}   & 11.47   & 11.84           & 16.33          & \multicolumn{1}{c|}{2135.47}    & 390.94             & --              \\
SCALE-LETKF & \textbf{11.59}   & 11.92   & 12.89           & 21.50          & \multicolumn{1}{c|}{2629.48}    & 351.61             & --              \\ \hline
\end{tabular}
}
\vspace{-2mm}
\caption{Compression ratio, compression time, and decompression time for each compressor with $\varepsilon = 0.04$ and error bound $\xi = 0.012$ (except TopoQZ has $e = \zeta = 0.006$).
Times are in seconds.
Trials that did not finish are marked with a dash. 
TopoQZ ran out of memory on Nyx and it crashed on Isabel and Tangaroa due to unknown reasons. 
TopoSZ ran out of memory on Nyx, S3D, and SCALE-LETKF. 
}
\label{tab:compression-task}

\includegraphics[width=\columnwidth]{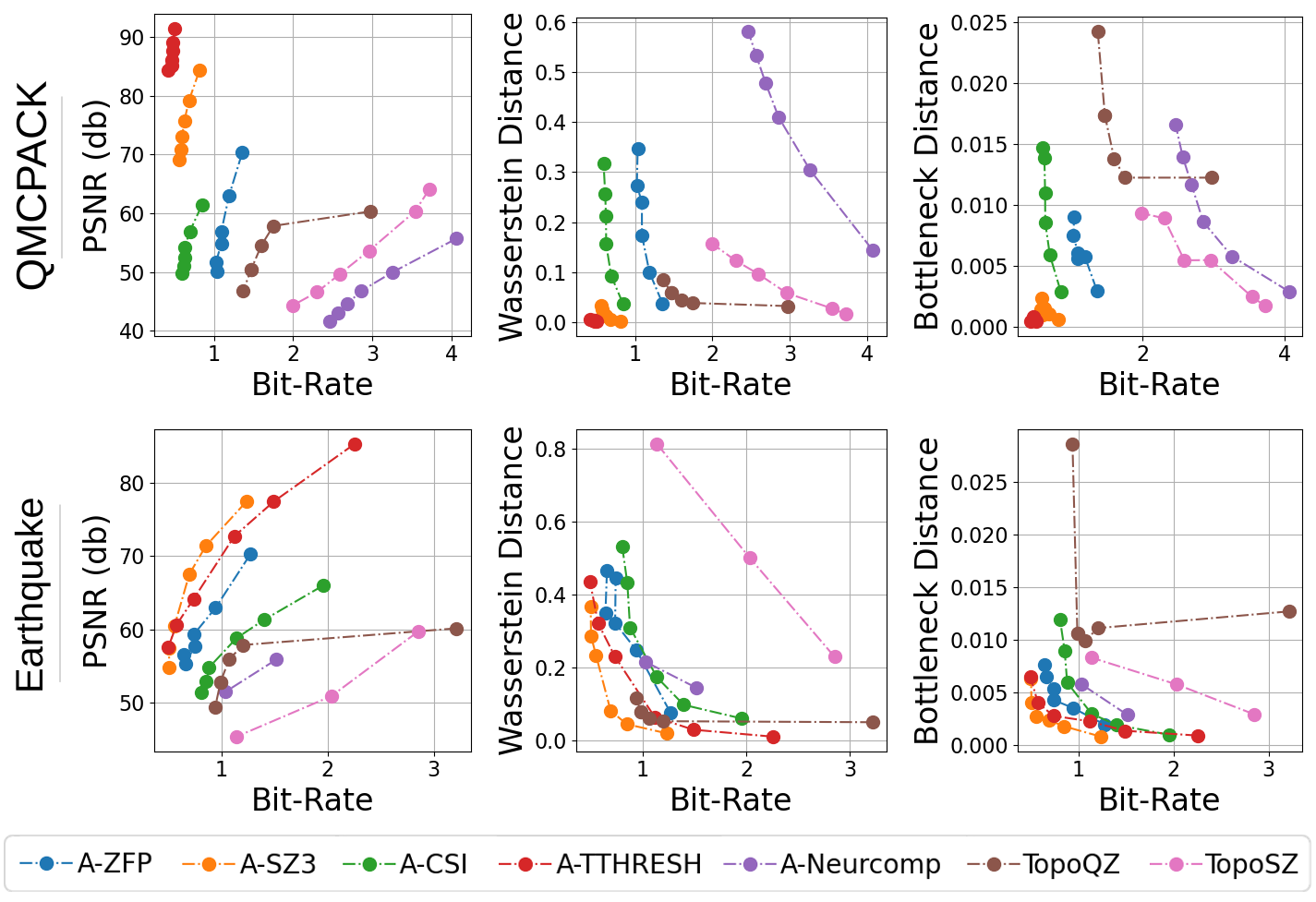}
    \vspace{-4mm}
    \captionof{figure}{PSNR, bottleneck distance, and Wasserstein distance versus bit-rate for each compressor for the QMCPACK and Earthquake datasets with $\varepsilon = 0.04$ ($e = 0.04$ for TopoQZ). These curves are given for other datasets in \cref{sec:reconstruction-quality-extra}.}
    \label{fig:reconstruction-quality}
    \vspace{-4mm}
\end{table}

%---------------------------------------
\subsubsection{Evaluation Metrics}
\label{sec:augmented-compressors-evaluation-metrics}

Compression ratio and times are reported in \cref{tab:compression-task} for a fixed parameter configuration of $\varepsilon = 0.04$ and $\xi = 0.012$.
We chose this parameter configuration because a small amount of persistence simplification preserves a large number of topological features in the input data, generating complex test cases for topology-preserving compression. For the reconstruction quality demonstrated in \cref{fig:reconstruction-quality}, $\varepsilon = 0.04$ is chosen similarly, and $\xi$ is varied between $0.003$ and $0.018$ to yield a variety of different compression ratios while still remaining small.
In this section, we compare the different augmented compressors. We leave the comparison with TopoQZ and TopoSZ to \cref{sec:compare-topology}.

\para{Compression ratios.} 
As shown in \cref{tab:compression-task}, Augmented TTHRESH produces the best compression ratios in every trial. Augmented Neurcomp performs noticeably worse than the other compressors.

\para{Reconstruction quality.}
In every trial, our framework successfully maintains a pointwise error bound $\xi$. There is a natural trade off between compression ratio and reconstruction quality. As shown in \cref{fig:reconstruction-quality}, Augmented SZ3 and Augmented TTHRESH have the best trade off between bit-rate, PSNR, $d_W$ and $d_B$, and perform equally well. Augmented Neurcomp performs the worst based on above metrics.

When visualized, we find that the decompressed volumes generally resemble the ground truth. However, when using certain transfer functions, visual artifacts may become visible. Artifacts appear more in visualizations that are sensitive to small changes in the transfer function. For the volume renderings in this paper, we chose transfer functions that led to fewer visual artifacts; see~\cref{sec:visual-artifacts} for adversarial examples.

\begin{figure}[!t]
    \begin{subfigure}[b]{0.06\linewidth}
        \raisebox{0.5\height}{\includegraphics[width=\linewidth]{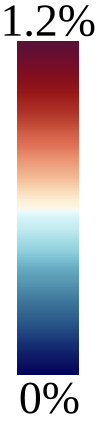}}
    \end{subfigure}
    \hfill
    \begin{subfigure}[b]{0.46\linewidth}
    \centering
        \includegraphics[width=\linewidth]{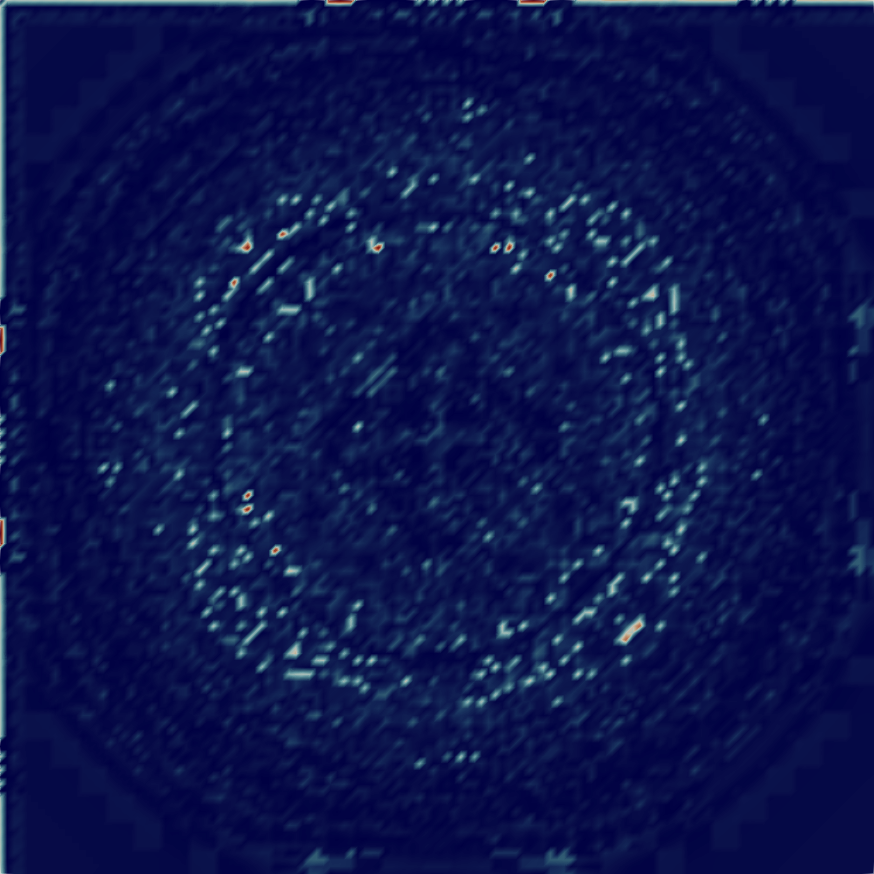}
        (A)
    \end{subfigure}
    \hfill
    \begin{subfigure}[b]{0.46\linewidth}
    \centering
        \includegraphics[width=\linewidth]{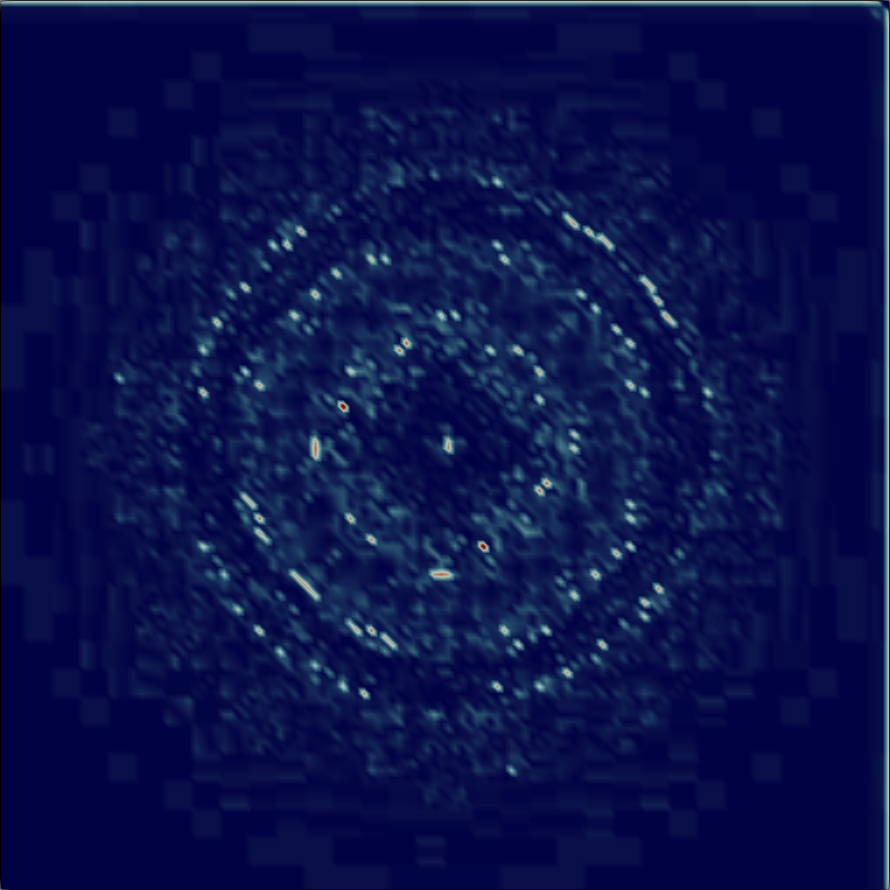}
        (B)
    \end{subfigure}
    \vspace{-5mm}
    \caption{Error map of a topologically complex slice of the Ionization dataset (A) before error bound tightening and (B) after error bound tightening.}
    \label{fig:errorMap}
    \vspace{-8mm}
\end{figure}

In practice, we find that upper and lower bound tightening does not affect PSNR very much; most of the reconstruction quality is determined by the initial upper and lower bounds. \cref{fig:errorMap} shows a map of the absolute error of each point for a topologically complex slice of the Ionization dataset before and after tightening. We can see that tightening does not have a significant effect on the average error.

\para{Run time analysis.} 
There are significant differences in run time among the augmented compressors.
As discussed in \cref{sec:run-time}, these times are affected by factors other than the base compression time. However, Augmented Neurcomp is the slowest because Neurcomp compresses data by training a neural network.  
Of the remaining four compressors, ZFP is typically the fastest, and TTHRESH the slowest, although this observation does not hold for all trials. For decompression time, ZFP is the fastest, while Neurcomp remains the slowest.

\para{Highlighted results.} 
There is no clear best augmented compressor that outperforms others across all metrics. Other than augmented Neurcomp, utilizing any augmented compressor plays a trade off between compression ability and speed. For the remainder of our analysis, we will primarily focus on Augmented ZFP, which is the fastest augmented compressor, and Augmented TTHRESH, which yields the best compression ratios and reconstruction quality.

%---------------------------------------
\subsection{Comparison with TopoQZ and TopoSZ}
\label{sec:compare-topology}

\para{Topological guarantees.}
Our framework preserves the contour tree during compression, and achieves the same topological guarantee as TopoSZ. TopoQZ ensures that all critical point pairs are preserved above a persistence threshold $\varepsilon$, but their locations and connectivity may be distorted after compression.

\para{Compression ratio.} 
In terms of compression ratio, when maintaining a strict topological constraint $\varepsilon = 0.04$ and error bound $\xi = 0.012$, every augmented compressor except Augmented Neurcomp outperforms both TopoQZ and TopoSZ in every trial.

\para{Reconstruction quality.} 
The curves in \cref{fig:reconstruction-quality} show that every augmented compressor except Augmented Neurcomp can match the PSNR of TopoQZ and TopoSZ,  while using less space. In terms of topological distance, the augmented compressors except Augmented Neurcomp outperform TopoSZ in terms of $d_W$ and $d_B$. They also outperform TopoQZ in terms of $d_B$, but are comparable in terms of $d_W$.

\para{Run time analysis.} 
In terms of compression time, the augmented compressors except Augmented Neurcomp produce times that are comparable to or better than TopoQZ, and significantly outperform TopoQZ on the largest datasets. 
These four augmented compressors are also significantly faster than TopoSZ across all trials. 

In terms of decompression time, the augmented compressors except Augmented Neurcomp perform slower than TopoSZ but faster than TopoQZ. There are several possible reasons why our decompression times are slower than TopoSZ. First and most notably, our decompression process is more complex, as it involves a decompression with the base compressor and then an augmentation of the decompressed results. This process requires more operations and has a higher I/O overhead. Second, we use XZ along with tar archives for lossless compression, which is slower than ZSTD used by TopoSZ. See \cref{sec:more-running-time} for a more detailed analysis of the decompression time.   

%---------------------------------------
\subsection{Analysis of Compression Time}
\label{sec:run-time}

\para{Asymptotic analysis.}
Let $n$ be the number of vertices in the rectilinear mesh. Our algorithm utilizes heap merges~\cite{gueunet2017task} during the merge tree computation; however, we use binary heaps (stored in arrays) instead of Fibonacci heaps from~\cite{gueunet2017task}. For a binary heap with $m$ elements, a single insertion operation has a worst-case time complexity of $O(\log m)$. 
Following~\cite{gueunet2017task}, from bottom to top, constructing an edge $e$ in a merge tree requires merging its heap with the heaps of its descendants, which takes $O(n \log n)$. Let $h$ denotes the \emph{height} of the tree, which corresponds to the maximum number of ancestor edges.
Then constructing a merge tree using these insertion-based heap merges takes $O(h n \log n)$. During the progressive tightening process, let $F$ denote the total number of detected false cases, each of which triggers a (partial) recomputation of the merge tree. Therefore, our algorithm takes $O(F h n \log n) = O(n^3 \log n)$. 

In practice, $F \ll n$ as shown in \cref{tab:time}. 
Additionally, $h \ll n$. We found that $\frac{h}{n}$ ranged from $0.0004$ (Miranda, A-ZFP) to $0.025$ (Nyx, A-Neurcomp). Excluding Augmented Neurcomp, $\frac{h}{n} < 0.01$ in $97\%$ of trials.  
On the other hand, using Fibonacci heaps to construct a merge tree~\cite{gueunet2017task} takes $O(n \log n)$ due to constant time heap merges; however, in our setting, we have found that binary heaps have lower run time in practice. Likewise, it is possible to merge heaps in linear time, but we instead merge by repeatedly inserting each element of the smaller heap into the larger one, as doing so has a much lower run time in practice.

\para{Empirical analysis.} To analyze the run time empirically, we calculate the amount of time for each portion of our algorithm with Augmented ZFP and Augmented TTHRESH,  with $\varepsilon = 0.04$ and $\xi = 0.012$. These run times are shown in \cref{tab:time}.

In \cref{tab:time}, the most time-consuming task is the computation of merge and contour trees. We compute the contour tree of the input data at the beginning of the algorithm. During the error bound tightening steps we also compute the contour tree of the decompressed data. These run times are shown in \cref{tab:time} under the `CT' and `Grow' columns, and account for $35-77\%$ of the total run time for each trial in \cref{tab:time}. 

For most of the trials, the time to run the base compressor, shown in the `BC' column, is a relatively small percentage of the overall compression time. 
However, if a base compressor produces results that nearly preserve the contour tree and does not produce too many extra branches, including those of persistence below $\varepsilon$, the augmentation time may be lower. This phenomenon suggests that the accuracy of the base compressor may have more effect on the total compression and augmentation time than just base compression.
In general, the run time of each base compressor is much faster than its augmented counterpart; see~\cref{sec:more-running-time} for a comparison. 

\begin{table}[!ht]
\setlength{\tabcolsep}{2pt}
\centering
\vspace{-2mm}
\resizebox{\columnwidth}{!}{
\begin{tabular}{cccccccccc}
\hline
\multicolumn{1}{c|}{Dataset}     & BC    & CT     & ULB     & Grow     & \%B    & \#FC & Fix      & File  & Total    \\ \hline
\multicolumn{10}{c}{Augmented ZFP}                                                                                  \\ \hline
\multicolumn{1}{c|}{QMCPack}     & 0.15  & 0.34   & 0.21   & 0.38     & 0.17\% & 0    & 0.0      & 0.27  & 1.35     \\
\multicolumn{1}{c|}{Tangaroa}    & 1.42  & 2.16   & 2.80   & 1.34     & 0.43\% & 14   & 0.0004   & 1.84  & 9.57     \\
\multicolumn{1}{c|}{Earthquake}  & 0.45  & 1.56   & 1.26   & 2.88     & 0.79\% & 2    & 0.0001   & 0.97  & 7.12     \\
\multicolumn{1}{c|}{Ionization}  & 0.70  & 1.56   & 2.09   & 3.10     & 1.15\% & 10   & 0.0009   & 1.35  & 8.81     \\
\multicolumn{1}{c|}{Isabel}      & 4.73  & 4.61   & 8.95   & 9.48     & 0.54\% & 1    & 0.0647   & 6.17  & 34.01    \\
\multicolumn{1}{c|}{Miranda}     & 4.00  & 156.79 & 17.92  & 37.85    & 1.21\% & 0    & 0.0      & 7.02  & 223.58   \\
\multicolumn{1}{c|}{NYX}         & 24.75 & 695.26 & 150.27 & 142.16   & 0.47\% & 6    & 4.60     & 42.21 & 1084.05  \\
\multicolumn{1}{c|}{S3D}         & 14.2  & 27.93  & 69.11  & 56.40    & 1.04\% & 39   & 0.01411  & 38.44 & 206.63   \\
\multicolumn{1}{c|}{SCALE-LETKF} & 15.2  & 16.15  & 61.67  & 92.99    & 0.63\% & 37   & 0.003408 & 35.34 & 221.47   \\ \hline
\multicolumn{10}{c}{Augmented TTHRESH}                                                                              \\ \hline
\multicolumn{1}{c|}{QMCPack}     & 0.23  & 0.34   & 0.21   & 0.45     & 0.77\% & 0    & 0.0      & 0.22  & 1.45     \\
\multicolumn{1}{c|}{Tangaroa}    & 3.34  & 2.2    & 2.79   & 1.92     & 0.77\% & 18   & 0.0001   & 1.59  & 11.84    \\
\multicolumn{1}{c|}{Earthquake}  & 1.53  & 1.56   & 1.25   & 3.33     & 0.65\% & 4    & 0.0001   & 0.91  & 8.58     \\
\multicolumn{1}{c|}{Ionization}  & 1.93  & 1.54   & 2.08   & 8.09     & 1.48\% & 5    & 0.0120   & 1.24  & 14.94    \\
\multicolumn{1}{c|}{Isabel}      & 13.13 & 4.62   & 9.01   & 10.24    & 0.54\% & 1    & 0.0666   & 5.62  & 42.69    \\
\multicolumn{1}{c|}{Miranda}     & 12.73 & 156.48 & 17.73  & 153.69   & 2.33\% & 0    & 0.0      & 7.24  & 347.87   \\
\multicolumn{1}{c|}{NYX}         & 57.87 & 687.03 & 132.49 & 15838.88 & 0.82\% & 89   & 97.82    & 37.52 & 27286.05 \\
\multicolumn{1}{c|}{S3D}         & 66.35 & 27.89  & 67.27  & 56.51    & 1.14\% & 47   & 0.0136   & 32.11 & 250.79   \\
\multicolumn{1}{c|}{SCALE-LETKF} & 69.21 & 16.10  & 60.00  & 193.18   & 0.54\% & 46   & 0.0032   & 29.71 & 368.35   \\ \hline
\end{tabular}
}
\vspace{-2mm}
\caption{Runtime analysis for each component of the augmented framework involving Augmented ZFP and Augmented TTHRESH with $\varepsilon = 0.04$ and $\xi = 0.012$. 
All times are in seconds. 
BC: running the base compressor.
CT: computing the contour tree of the input data.
ULB: calculating the initial upper and lower bounds. 
Grow: time growing the contour tree of the reconstructed data.
\%B: percent of branches in the reconstructed contour tree whose growth was recomputed.
\#FC: number of false cases corrected after upper and lower bounds are set.
Fix: average time to fix a false case, excluding regrowing branches.
File: average time to write the compressed output to a file.}
\label{tab:time}
\vspace{-6mm}
\end{table}
\section{Conclusion and Discussion}
\label{sec:conclusion}

We introduce a novel framework for augmenting \emph{any} lossy compressor to preserve the contour tree of a volumetric dataset while maintaining a user-specified global error bound. 
To do this, our framework first imposes topology-informed upper and lower bounds on each data point. 
It then progressively tightens those bounds until the contour tree is preserved. 
We also introduce a novel encoding scheme that efficiently stores individual points with variable precision and maintains these upper and lower bounds. 
When our framework is used to augment state-of-the-art lossy compressors, it is shown to preserve the contour trees of various scientific datasets.
Our augmented compressors also achieve higher compression ratios and reconstruction quality than those obtained by existing topology-preserving compressors in comparable or faster time.
Our framework will benefit from any advancement with lossy compression since it can be used to augment increasingly effective lossy compressors to achieve better topology-preserving compression. 

Our framework is not without limitations. The compression times are longer than the base compressors. This difference gets worse as the topological complexity of the data increases.
However, in some use-cases, topological preservation is preferable to run time.
Regardless, our framework would benefit from more efficient or parallel implementations for the contour/merge tree computation and the encoding scheme.

%-------------------------
\acknowledgments{
This work was supported in part by grants from National Science Foundation OAC-2330367, OAC-2313122, 
OAC-2313123, OAC-2311878,
and OAC-2313124. 
}
%-------------------------

\bibliographystyle{abbrv-doi-hyperref}
\bibliography{refs-compression.bib}

%-------------------------
\clearpage
\newpage
\pagenumbering{arabic}
\renewcommand*{\thepage}{A\arabic{page}}
\appendix 
\section{Details on the Datasets}
\label{sec:datasets}

We include details on each dataset used in our experiments. Some datasets are accessed from the SDR Bench \cite{zhao2020sdrbench}, available at~\cite{SDRBench}. Datasets accessed from the SDR Bench may involve contributions from the DOE NNSA ECP project and the ECP CODAR project.

The \textbf{QMCPACK} dataset (accessed from the SDR Bench) comes from the QMCPACK performance test using the QMCPACK continuum quantum Monte Carlo
simulation \cite{Kim_2018, Kent2020}. Only the 145th orbital out of 288 is used for testing, for which only a single field is provided. The dataset is normalized to $[0,1]$ before compression. 

The \textbf{Tangaroa} dataset comes from a single time frame simulating the wind flow around a 3D model of the Research Vessel Tangaroa \cite{popinet2004experimental}. 
The magnitude of the wind velocity is used as the scalar field of interest. The dataset is normalized to $[0,1]$ before compression.

The \textbf{Earthquake} dataset originates from a TeraShake 2 earthquake simulation~\cite{olsen2008terashake2} and has been part of the 2006 IEEE Visualization Design Contest~\cite{scivis2006}.  
The dataset used in this paper is obtained from the public data repository of Pont et al. used for their publication~\cite{pont2021wasserstein}. The dataset is preprocessed by Pont et al. and comes with a single field. It represents one time step of a simulation of an earthquake at the San Andreas fault. Specifically, we use time step 011700. The details of the preprocessing can be found on the repository. The dataset is normalized to $[0,1]$ before compression.

The \textbf{Ionization} dataset originates from an ionization front simulation by Whalen and Norman \cite{whalen2008ionization} and has been featured in the 2008 IEEE Visualization Design Contest~\cite{scivis2008}.   
The simulation is done with 3D radiation hydro-dynamical calculations of ionization front instabilities in which multi-frequency radiative transfer  is coupled to the primordial chemistry of eight species~\cite{whalen2008ionization}. 
The single time step used in this paper comes from cluster 2, time step 0125 and is obtained from the same repository as the Earthquake dataset. It is 
preprocessed and comes with a single field. The details of the preprocessing can be found on the repository. The dataset is normalized to $[0,1]$ before compression.

The \textbf{Isabel} dataset originates from a hurricane simulation from the National Center for Atmospheric Research, and has been included in the 2004 IEEE Visualization Design Contest \cite{scivis2004}.
While the original dataset has a size of  $500 \times 500 \times 100$, we truncate the dataset to $500 \times 500 \times 90$ in order to avoid land regions that contain no data values. We use the wind speed field. The dataset is normalized to $[0,1]$ before compression.

The \textbf{Miranda} dataset (accessed from the SDR Bench) comes from the hydrodynamics code for large turbulence simulations conducted by Lawrence Livermore National Laboratory. We use the density field.

The $\textbf{Nyx}$ dataset (accessed from the SDR Bench) comes from the Nyx cosmological simulation~\cite{almgren2013nyx}. We use the dark matter density field.

The $\textbf{S3D}$ dataset comes from the S3D turbulence simulation software~\cite{treichler2017s3d}. This dataset is derived from data from the SDR Bench. We compute the field as the magnitude of the velocity, as derived from the velocity $x$, $y$, and $z$ components provided. We use its parameter setting $1.7e \times 10^{-3}$.

The $\textbf{SCALE-LETKF}$ dataset (accessed from the SDR Bench) comes from the Local Ensemble Transform Kalman Filter (LETKF) data assimilation package for the SCALE-RM weather model~\cite{lien2017near}. We use the QV field which is up-sampled to double precision.
\section{Evaluation Metrics}
\label{sec:evaluationMetrics}

We evaluate our trials on several metrics. We measure compression ratio, compression time, and peak signal-to-noise ratio (PSNR), which are common metrics for evaluating compressors. 

The compression ratio is the size of the uncompressed file divided by the size of the compressed file, and higher compression ratios indicate smaller compressed file sizes. Related to this, the bit-rate is the number of bits used to compress a single data point. For floating point data, it is either equal to 32 divided by the compression ratio, or 64 divided by the compression ratio, depending on if the dataset is respectively single or double precision.

Peak signal-to-noise ratio (PSNR) is a number that measures reconstruction quality. If $R$ is the range of the data and $M$ is the mean squared error of our reconstruction, the PSNR is defined as:
\begin{equation}
  \label{eq:psnr}
  \text{PSNR} = 10 \log_{10}\left( \frac{ R^2 }{ M } \right).
\end{equation}
In general, higher PSNR values indicate higher reconstruction quality. We also measure the bottleneck distance \cite{cohen2005stability} and Wasserstein distance \cite[page 183]{edelsbrunner2022computational} to quantify the amount of topological control.

To define the Wasserstein and bottleneck distances, suppose that $f$ and $f'$ are two scalar fields that yield persistence diagrams $D$ and $D'$ respectively; see~\cite{EdelsbrunnerLetscherZomorodian2002} for an introduction to persistent homology. If $\varphi:D \rightarrow D'$ represents any bijection, and $q \geq 1$, we define the $q$-Wasserstein distance $W_q$ as:
\begin{equation}
W_q(D,D') = \left[ \inf_{\varphi:D\rightarrow D'} \sum_{p \in D} || p - \varphi(p)||_\infty^q \right]^{\frac{1}{q}}.
\end{equation}
For our evaluation, we set $d_W = W_2$ and $d_B = W_\infty$. In particular:
\begin{equation}
W_\infty(D,D') = \inf_{\varphi:D \rightarrow W'} \sup_{p \in D} ||p- \varphi(p)||_\infty. 
\end{equation}

In general, lower values of $W_q$ for any $q \geq 1$ indicate that the persistence diagrams $D$ and $D'$ are more similar. This in-turn means that the datasets that produced $D$ and $D'$ are more topologically similar. To compute $d_W$ and $d_B$ in reasonable time, we report approximate values~\cite{KerberMorozovNigmetov2016} with $1\%$ error that were computed after persistence simplification with a threshold of $1.5 \times 10^{-6}$ times the range of the data.
\section{Analysis of Compression and Decompression Times}
\label{sec:more-running-time}

We provide compression times for the base compressor versus the augmented compressors in~\cref{sec:base-compressor-times}. We provide a detailed analysis of decompression times in~\cref{sec:decompression-time-analysis}.

\subsection{Compression Times}
\label{sec:base-compressor-times}

We give the run times (in seconds) of each base compressor versus the augmented compressor in~\cref{tab:base-compressor-times}. Each augmented compressor is abbreviated as A-ZFP, A-SZ3, etc. 
Times are for $\varepsilon = 0.04$ and $\xi = 0.012$. The parameter settings for each base compressor are the same as those used prior to the augmentation (see~\cref{sec:base-compressor-parameters} for specifics).

\begin{table*}[!ht]
\scriptsize
\centering{
\begin{tabular}{l|cc|cc|cc|cc|cc}
\hline
Dataset & \multicolumn{1}{l}{ZFP} & \multicolumn{1}{l|}{A-ZFP} & \multicolumn{1}{l}{SZ3} & \multicolumn{1}{l|}{A-SZ3} & \multicolumn{1}{l}{CSI} & \multicolumn{1}{l|}{A-CSI} & \multicolumn{1}{l}{TTHRESH} & \multicolumn{1}{l|}{A-TTHRESH} & \multicolumn{1}{l}{Neurcomp} & \multicolumn{1}{l}{A-Neurcomp} \\ \hline
QMCPACK     & 0.04                    & 1.3                        & 0.05                    & 1.37                       & 0.03                    & 1.21                       & 0.16                        & 1.49                           & 1633.42                      & 1640.44                        \\
Tangaroa    & 0.24                    & 9.79                       & 0.34                    & 11.26                      & 0.99                    & 9.46                       & 1.69                        & 12.11                          & 26.20                        & 63.17                          \\
Earthquake  & 0.35                    & 7.42                       & 0.24                    & 7.47                       & 0.14                    & 7.16                       & 1.16                        & 8.86                           & 1021.93                      & 1042.63                        \\
Ionization  & 0.24                    & 8.96                       & 0.34                    & 10.24                      & 0.20                    & 14.31                      & 1.27                        & 15.29                          & 1195.47                      & 1222.28                        \\
Isabel      & 0.73                    & 34.85                      & 1.00                    & 36.30                      & 3.47                    & 43.53                      & 7.73                        & 43.30                          & 7009.20                      & 7151.38                        \\
Miranda     & 1.31                    & 225.75                     & 2.25                    & 285.79                     & 1.36                    & 250.63                     & 8.61                        & 353.24                         & 8865.54                      & 9377.60                        \\
S3D         & 5.28                    & 214.86                     & 7.27                    & 258.83                     & 5.25                    & 178.04                     & 53.10                       & 258.56                         & 31976.49                     & 34663.49                       \\
Nyx         & 1.78                    & 1072.75                    & 6.14                    & 1147.18                    & 21.46                   & 5716.34                    & 31.74                       & 25596.02                       &   87230.00                           &       29991.31                         \\
SCALE-LETKF & 4.99                    & 226.58                     & 8.01                    & 402.86                     & 5.73                    & 347.17                     & 49.83                       & 375.18                         & 37072.43                     & 41227.85                       \\ \hline
\end{tabular}}
\vspace{-2mm}
\caption{Compression times (in seconds) for each base compressor versus each augmented compressor on each dataset. Augmented compressors are abbreviated as A-ZFP, A-SZ3, etc.}
\label{tab:base-compressor-times}
\vspace{-4mm}
\end{table*}

\subsection{Decompression Time}
\label{sec:decompression-time-analysis}

We provide a breakdown of the decompression times into multiple stages for the Augmented ZFP and Augmented TTHRESH on each dataset in \cref{tab:decompression-time}. 
For comparison, the decompression times for TopoSZ are as follows: QMCPACK: $0.01$s; Tangaroa: $0.12$s; Earthquake: $0.07$s; Ionization: $0.10$s; Isabel: $0.41$s; and Miranda: $0.67$s.

The decompression pipeline of an augmented compressor first decompresses a compressed file using XZ with a tar archive, and uses Huffman to decode the quantization numbers. Then, it decompresses the compressed file using the base decompressor (run as a separate process), and augments the result using the quantization numbers. Finally, it saves the output. As shown in~\cref{tab:decompression-time}, no individual stage of the decompression pipeline dominates the others in terms of run time, although running the base decompressor and augmentation takes  longer times. 

\para{Comparison to TopoSZ.} Our decompression times (using augmented compressors) are much slower than TopoSZ for several reasons. First, the decompression pipeline of TopoSZ does not involve any augmentation;  whereas our pipeline runs a base decompressor and augments the result. Augmentation increases the I/O overhead, as our augmentation layer must read in the output of the base decompressor, which is much larger than the single compressed file read in by TopoSZ. Second, for lossless compression, we decompress using XZ, whereas TopoSZ uses ZSTD, which is much faster; see \cref{tab:XZ-versus-ZSTD} for a comparison. Furthermore, because we must store the output of the base decompressor and the augmentation layer, we must combine both files into a tar archive. Extracting this tar archive takes additional time. Finally, the code for our decompression pipeline is not optimized and has room for a more efficient implementation. 

\begin{table}[!ht]
\setlength{\tabcolsep}{2pt}
\centering
\begin{tabular}{cccccccc}
\hline
\multicolumn{1}{c|}{Dataset}     & Lossless & Load & Base  & Augment & Save & Clean & Total \\ \hline
\multicolumn{7}{c}{Augmented ZFP}                                                   &       \\ \hline
\multicolumn{1}{c|}{QMCPACK}     & 0.03     & 0.01 & 0.03  & 0.01    & 0.01 & 0.04  & 0.14  \\
\multicolumn{1}{c|}{Tangaroa}    & 0.11     & 0.07 & 0.10  & 0.11    & 0.04 & 0.05  & 0.48  \\
\multicolumn{1}{c|}{Earthquake}  & 0.06     & 0.07 & 0.11  & 0.06    & 0.05 & 0.04  & 0.39  \\
\multicolumn{1}{c|}{Ionization}  & 0.08     & 0.10 & 0.14  & 0.08    & 0.06 & 0.04  & 0.5   \\
\multicolumn{1}{c|}{Isabel}      & 0.26     & 0.21 & 0.32  & 0.36    & 0.12 & 0.05  & 1.33  \\
\multicolumn{1}{c|}{Miranda}     & 0.27     & 0.69 & 0.76  & 0.62    & 0.36 & 0.07  & 2.78  \\
\multicolumn{1}{c|}{Nyx}         & 1.18     & 1.38 & 2.12  & 2.37    & 1.38 & 0.09  & 8.52  \\
\multicolumn{1}{c|}{S3D}         & 2.22     & 2.38 & 2.70  & 2.73    & 1.24 & 0.15  & 11.42 \\
\multicolumn{1}{c|}{SCALE-LETKF} & 1.57     & 2.66 & 2.90  & 2.54    & 1.45 & 0.14  & 11.26 \\ \hline
\multicolumn{7}{c}{Augmented TTHRESH}                                               &       \\ \hline
\multicolumn{1}{c|}{QMCPACK}     & 0.03     & 0.02 & 0.06  & 0.01    & 0.01 & 0.04  & 0.16  \\
\multicolumn{1}{c|}{Tangaroa}    & 0.08     & 0.07 & 0.58  & 0.11    & 0.05 & 0.05  & 0.94  \\
\multicolumn{1}{c|}{Earthquake}  & 0.05     & 0.07 & 0.30  & 0.06    & 0.05 & 0.05  & 0.58  \\
\multicolumn{1}{c|}{Ionization}  & 0.06     & 0.11 & 0.48  & 0.09    & 0.07 & 0.05  & 0.85  \\
\multicolumn{1}{c|}{Isabel}      & 0.10     & 0.23 & 1.77  & 0.36    & 0.13 & 0.05  & 2.64  \\
\multicolumn{1}{c|}{Miranda}     & 0.21     & 0.72 & 2.47  & 0.65    & 0.38 & 0.08  & 4.5   \\
\multicolumn{1}{c|}{Nyx}         & 1.02     & 1.37 & 4.21  & 2.28    & 0.72 & 0.10  & 9.7   \\
\multicolumn{1}{c|}{S3D}         & 1.48     & 2.46 & 8.61  & 2.54    & 1.36 & 0.14  & 16.59 \\
\multicolumn{1}{c|}{SCALE-LETKF} & 0.96     & 2.73 & 13.57 & 2.49    & 1.48 & 0.15  & 21.39 \\ \hline
\end{tabular}
\vspace{-2mm}
\caption{Run times (in seconds) associated with the decompression pipeline for Augmented ZFP and Augmented TTHRESH with $\varepsilon = 0.04$ and $\xi = 0.012$.  
Lossless: time to losslessly decompress with XZ, tar archive, and Huffman coding.
Load: time spent loading data from files.
Base: time to decompress using the base decompressor.
Augment: time spent augmenting the output of the base decompressor.
Save: time spent saving the result to a RAW binary file.
Clean: time spent deleting files using system calls.}
\label{tab:decompression-time}
\end{table}

\begin{table}[!ht]
\setlength{\tabcolsep}{2pt}
\centering
%\resizebox{\columnwidth}{!}{
\begin{tabular}{c|cc|cc}
\hline
Dataset     & XZ               & ZSTD             & XZ                & ZSTD              \\ \hline
Compressor  & \multicolumn{2}{c|}{A-ZFP} & \multicolumn{2}{c}{A-TTHRESH} \\ \hline
QMCPACK     & 0.02             & 0.02             & 0.02              & 0.01              \\
Earthquake  & 0.04             & 0.02             & 0.03              & 0.02              \\
Ionization  & 0.04             & 0.02             & 0.03              & 0.02              \\
Tangaroa    & 0.07             & 0.02             & 0.05              & 0.03              \\
Isabel      & 0.21             & 0.04             & 0.07              & 0.03              \\
Miranda     & 0.18             & 0.05             & 0.12              & 0.05              \\
SCALE-LETKF & 1.33             & 0.15             & 0.76              & 0.11              \\
S3D         & 1.94             & 0.16             & 1.28              & 0.11              \\
Nyx         & 0.78             & 0.12             & 0.65              & 0.11              \\ \hline
\end{tabular}
%}
\vspace{-2mm}
\caption{Lossless decompression time (in seconds) for the output of augmented ZFP (listed as A-ZFP) and augmented TTHRESH (listed as A-TTHRESH) for each dataset using XZ versus ZSTD.}
\label{tab:XZ-versus-ZSTD}
\vspace{-6mm}
\end{table}
\section{Compressor Parameter Settings}
\label{sec:base-compressor-parameters}

We discuss the parameter settings associated with base and augmented compressors. 
The parameter settings during augmentations are described in~\cref{sec:base-parameters-during-augmentation}. The parameter settings for ZFP and TTHRESH used to generate \cref{fig:teaser}, \cref{fig:zoom}, and \cref{tab:base-false-cases} are given in~\cref{sec:base-parameters-equivalent}.

\subsection{Parameter Settings During Augmentation}
\label{sec:base-parameters-during-augmentation}

When augmenting ZFP, SZ3, and TTHRESH, to preserve a pointwise error bound $\xi$, we set the error bound $\delta$ of ZFP to be $5\xi$, the error bound $\eta$ of SZ3 to be $0.25\xi$, and the target PSNR $\tau$ of TTHRESH to be $0.05\xi$. 
It seems intuitive that setting $\delta = \eta = \tau = \xi$ would be optimal. We provide a brief justification for why our strategy works better.

In the case of TTHRESH and SZ3, these parameter settings cause the base compressors to preserve data with higher reconstruction quality. If the base compressor returns higher quality compression, then our topology-preserving framework will need to make fewer adjustments in order to preserve the contour tree. 
We have found that the extra storage space used by the base compressors to create higher quality intermediate data is outweighed by the space saved by storing fewer adjustments, causing compression ratios to increase. In addition to improved compression ratios, if the intermediate data more closely resembles the input (ground truth) data, there will be fewer false cases, leading to lower augmentation time. Further, if the intermediate data is closer to the input data, then after augmentation, the decompressed data will also be closer to the input data, leading to a higher PSNR.

ZFP is very conservative when it comes to error bounds and actually preserves an error bound much lower than the maximum error bound set by the user. We notice empirically that by multiplying the maximum error bound for ZFP by five, this raises the actual error bound preserved by ZFP up to something closer to $\xi$, allowing for ZFP to produce higher compression ratios while still yielding intermediate data of acceptable reconstruction quality. 
Surprisingly, we observe that setting $\delta$ very high, such as $\delta = 50\xi$ yields even higher compression ratios than $\delta = 5\xi$. On the other hand, the compression ratios are not too high such that ZFP obtains a comparative advantage in terms of compression ratio, and the compression times are slow enough that ZFP loses its comparative advantage on time efficiency. Therefore, we determine that $\delta = 5\xi$ is the best parameter setting for ZFP.

\subsection{Parameters for Equivalent Compression Ratios}
\label{sec:base-parameters-equivalent}

In order to generate \cref{fig:teaser}, \cref{fig:zoom}, and \cref{tab:base-false-cases}, we need to identify parameter settings for the absolute error bound parameter $\delta$ of ZFP and the RMSE target parameter $\tau$ for TTHRESH which produce compression ratios comparable to Augmented ZFP and 
Augmented TTHRESH with $\epsilon = 0.04$ and $\xi = 0.012$. The values for $\delta$ and $\tau$ that produce these comparable compression ratios are reported in~\cref{tab:base-false-cases}.

\begin{table}[!h]
\centering
\begin{tabular}{l|ll}
\hline
Dataset    & ZFP $\delta$ & TTHRESH $\tau$ \\ \hline
QMCPACK    & 3.11e-2    & 7.96e-6        \\ \hline
Tangaroa   & 3.60e-2    & 2.15e-4        \\ \hline
Earthquake & 3.60e-2    & 3.66e-4        \\ \hline
Ionization & 4.98e-2    & 3.51e-4        \\ \hline
Isabel     & 1.25e-1    & 2.62e-4        \\ \hline
Miranda    & 2.28e-1    & 6.66e-5        \\ \hline
Nyx        & 1.97e-3    & 1.97e-3        \\ \hline
S3D        & 3.08e-5    & 3.43e-7        \\ \hline
SCALE-LETKF& 9.00e-4    & 2.95e-6        \\ \hline
\end{tabular}
\vspace{-2mm}
\caption{Parameter configurations used to obtain \cref{fig:teaser}, \cref{fig:zoom}, and \cref{tab:base-false-cases}.}
\vspace{-2mm}
\end{table}
In some cases, the parameter configuration for ZFP is higher than $0.012$. ZFP is very conservative in terms of error bounds, and for each of those trials, the actual error bound that is maintained is likely significantly lower than the stated maximum error bound.
\section{Algorithmic Details}
\label{sec:algorithm-details}

We describe how the initial upper and lower bounds are computed in~\cref{sec:set-bounds}. 
We provide an illustration for the false case detection process in~\cref{sec:false-case-figures}, followed by special cases in~\cref{sec:special-cases}. We describe the specifics surrounding the tightening process, including a comparison with TopoSZ in~\cref{sec:tighten-bounds}.

\subsection{Initial Upper and Lower Bounds}
\label{sec:set-bounds}

\begin{figure}[!ht]
\centering
\includegraphics[width=0.6\linewidth]{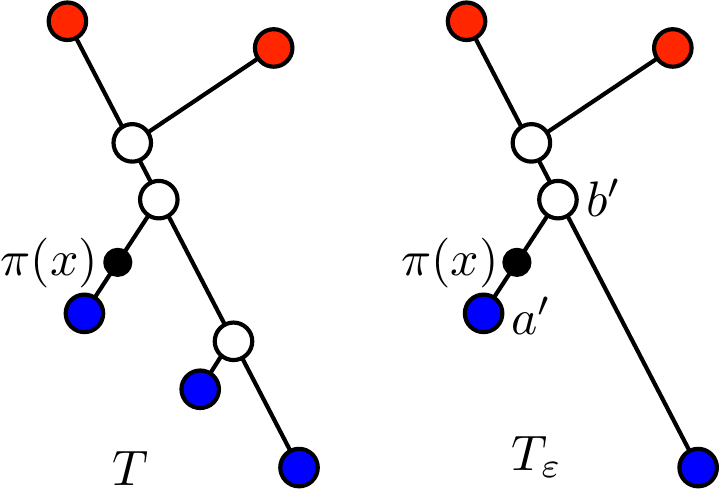}
\vspace{-2mm}
\caption{An illustration of Case A.}
\label{fig:upper-lower-a}
\vspace{-2mm}
\end{figure}

To compute the initial upper and lower bounds $U(x)$ and $L(x)$ for $x \in \X$, we aim to locate an edge $ab$ of the simplified contour tree $T_{\varepsilon}$ satisfying $f(a) \leq f(x) \leq (b)$. Recall that a contour tree $T$ of the data $(\X,f)$ arises from a quotient map $\pi: (\X, f) \to (\X/{\sim}, f)$. The contour tree $T_{\varepsilon}$ is defined analogously. With an abuse of notation, we use $\pi(x)$ to represent the image of $x$ under a quotient map $\pi$ (when $\pi$ is clear from the context). For instance, in \cref{fig:upper-lower-a}, $\pi(x)$ on the left arises from the quotient map defining $T$, whereas $\pi(x)$ on the right comes from the quotient map defining $T_{\varepsilon}$.   

\underline{Case A:}~As illustrated in \cref{fig:upper-lower-a}, let $e'=a'b'$ be an edge of $T_{\varepsilon}$ whose segmentation region contains $x$; that is, $\pi(x) \in e'$.  
If $f(a') \leq f(x) \leq f(b')$, then we set $ab := a'b'$. Otherwise, we need to locate $ab$ with some care, following Case B below. 

\underline{Case B:}~We begin with the join and split trees of $f$, denoted as $J$ and $S$, respectively. Let $J_\varepsilon$, $S_\varepsilon$, and $T_\varepsilon$ denote the joint, split, and contour  trees of $f_\varepsilon$, respectively. $J_{\varepsilon}$ and $S_{\varepsilon}$ combine to form the simplified contour tree $T_{\varepsilon}$. In Case B, if an edge in $J$ that contains $\pi(x)$ was removed after persistent simplification, then $\pi(x)$ maps to the saddle of the removed branch. We first compute edges $e'_J$ and $e'_S$ from $J_{\varepsilon}$ and $S_{\varepsilon}$ respectively as follows.    

To compute $e'_J$, we first find an edge $e_J$ of $J_\varepsilon$ whose segmentation region contains $x$. 
That is, $\pi(x) \in e_J$. Suppose that $e_J$ has endpoints $j_1$ and $j_2$. We then consider two cases for $x$.

\begin{figure}[!ht]
\centering
\includegraphics[width=0.6\linewidth]{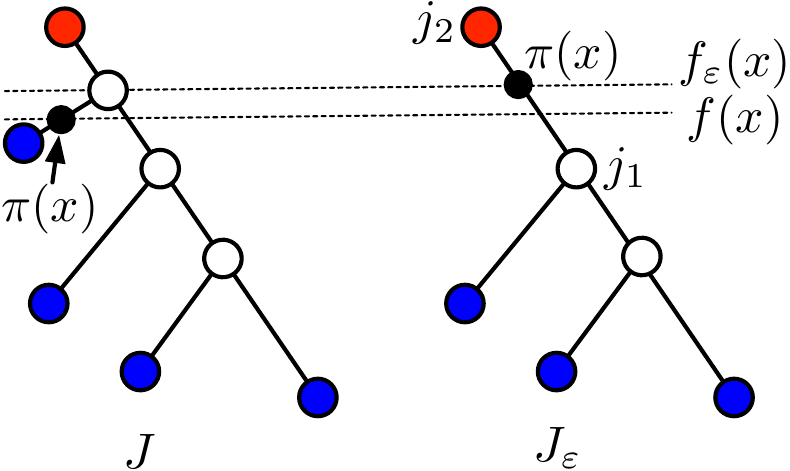}
\vspace{-2mm}
\caption{An illustration of Case B.1.}
\label{fig:upper-lower-b-1}
\vspace{-2mm}
\end{figure}

\underline{Case B.1:} $f(j_1) < f(x) < f(j_2)$. In this case, we set $e_J' = e_J$. As shown in~\cref{fig:upper-lower-b-1}, $x$ belongs to the segmentation region of edge $j_1j_2$ in $J_{\varepsilon}$, i.e., $\pi(x) \in j_1j_2$, and $f(j_1) < f(x) < f(j_2)$; so we set $e_J' = e_J = j_1j_2$.

\begin{figure}[!ht]
\centering
\includegraphics[width=0.6\linewidth]{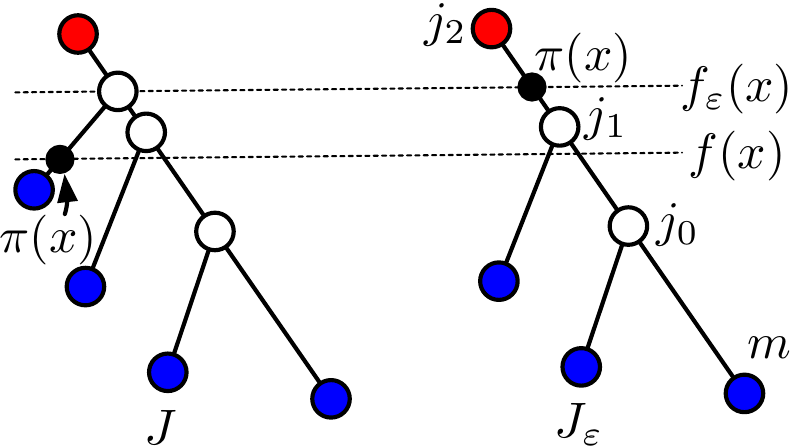}
\vspace{-2mm}
\caption{An illustration of Case B.2.}
\label{fig:upper-lower-b-2}
\vspace{-2mm}
\end{figure}

\underline{Case B.2:} $f(x) < f(j_1)$. In this case, let $m$ be the lowest leaf that descends from $e_J$ in $J_\varepsilon$. Then there exists a unique path connecting $j_1$ with $m$ that contains a sequence of descendant edges $e_1, e_2, \ldots$, and so on. There exists exactly one edge $e_k$ with endpoints $a_k$ and $b_k$ such that $f(a_k) < f(x) < f(b_k)$. We set $e_J' = e_k$. As shown in~\cref{fig:upper-lower-b-2}, we see that $x$ belongs to the segmentation region for edge $j_1j_2$ in $T_{\varepsilon}$, but $f(j_1) > f(x)$. We identify the edge $e_k = j_0j_1$ such that $f(j_0) < f(x) < f(j_1)$. Thus, we instead set $e_J' = j_0j_1$. 

We compute an edge $e'_S$ from $S_{\varepsilon}$ similarly to $e'_J$. Let $R'_J$ and $R'_S$ be the segmentation region of $e'_J$ and $e'_S$ respectively. 
Then $R'_J \cap R'_S$ will intersect the segmentation region of a single edge $ab$ of $T_{\varepsilon}$, and edge $ab$  will satisfy $f(a) < f(x) < f(b)$. 
We then set $L(x) = f(a) + \zeta$ and $U(x) = f(b) - \zeta$, where $\zeta = 10^{-5}|f(a)-f(b)|$, as described in \cref{sec:method-overview}.

\subsection{False Case Detection}
\label{sec:false-case-figures}

\begin{figure}[!ht]
\centering
\includegraphics[width=0.8\linewidth]{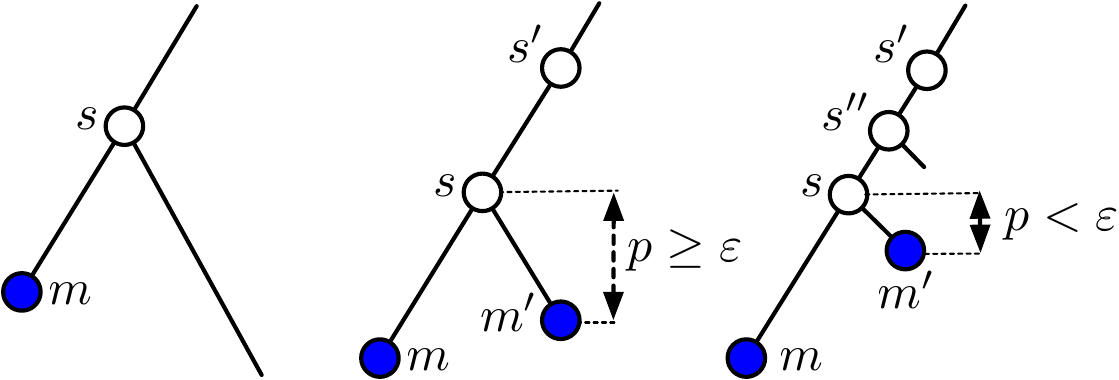}
\caption{From left to right: illustrations of Cases (I), (II.a) and (II.b) of the false case detection process respectively.}
\label{fig:false-case-figures}
\end{figure}

In~\cref{sec:augment-tightening}, we introduce false case detection as part of the progressive upper and lower bound tightening process. 
We illustrate Case (I), Case (II.a) and Case (II.b) respectively in~\cref{fig:false-case-figures}. 

We grow a minimum $m$ until it reaches some saddle $s$. 
In Case (I), we handle the situation where $s$ is unpaired. 
In Case (II), we handle the situation where $s$ has already been paired with another local minimum $m'$. As a result, $m$ must pair with some other saddle $s'$ with $f(s') > f(s)$.

In Case (II.a), $|f(m') - f(s)| \geq \varepsilon$. It follows that $|f(m) - f(s')| \geq \varepsilon$. As a result, we know that $m$ has $s$ as its parent in the simplified join tree.
In case (II.b), $|f(m') - f(s)| \leq \varepsilon$. Thus, the edge $m's$ is not part of the simplified join tree. Instead, we search for some new saddle $s''$ that is the parent of $s$. Then, we can check the cases (I) and (II) again.

\subsection{Special Cases}
\label{sec:special-cases}

There are a number of special cases that must be handled during upper and lower bound tightening. Let $f$ represent the ground truth scalar field. Let $f'$ be the decompressed scalar field. Recall that the upper and lower bound tightening step works on computing the contour tree of $f'$. Let $T$ be the merge tree of $f$, and let $T_\varepsilon$ be $T$ after persistence simplification.
Let $T'$ and $T'_\varepsilon$ be defined analogously for $f'$.

When applicable, assume that we are working with the join tree. All cases apply analogously for the split tree.

\para{Simulation of Simplicity.} 
It is possible that points may have the same scalar field value, making the function non-Morse. To resolve this issue, we use Simulation of Simplicity~\cite{edelsbrunner1990simulation}. If points $x$ and $y$ have the same scalar field value, we break the tie by treating $x$ as larger if its value is stored in memory after that of $y$ (assuming a Fortran order). %Gueunet et al. used the same resolution in their merge tree algorithm~\cite{gueunet2017task}.

\para{Higher-order saddles.} 
In a typical merge tree, each interior node (saddle) is connected to exactly two local minima. However, it is possible to have monkey saddles (i.e., saddles that are connected to three local minima) and higher-order saddles (i.e., saddles with more than three local minima connections).

%has exactly two children. However, it is possible that an interior node can have three or more children.

%, and multiple children are paired with that saddle for persistence. This can lead to difficulties with our algorithm. For example, suppose that a saddle $s$ has three children (local minima) $m_1$, $m_2$ and $m_3$ with $f(m_1) > f(m_2) > f(m_3)$. In this degenerate case, $m_1$ and $m_2$ are both paired with saddle $s$ for persistence. If the error bound tightening step were executed precisely as is written in \cref{sec:augment-tightening}, $m_1$ would be grown before $m_2$ and $m_3$ because it has the largest scalar field value. When $m_1$ is grown, it would pair with $s$. After this, $m_2$ would be grown, but $m_2$ would not yet be paired with $s$. This leads to a mis-pairing of critical points which would lead to compression errors.

To obtain the correct pairings involving higher-order saddles, we proceed as follows. 
Let $s$ be an interior saddle point. 
Let $N$ be the set of neighbors of saddle $s$. 
Let $L(s) = \{x \in N : f'(x) < f'(s)\}$ be the \textit{lower link} of $s$. When the growth of a local minimum $m$ terminates at $s$, we verify whether each point in the lower link $L(s)$ has already been visited during the growth of some child of $s$ (including possibly $m$). If every point in $L$ has not yet been visited, then we pair $m$ with $s$. Otherwise, $m$ must pair with some other saddle $s'$. This strategy is adapted from the merge tree algorithm of Gueunet et al.~\cite{gueunet2017task}.

\para{Out-of-order growth.} Let $s$ be an interior saddle point. Let $m_1$ and $m_2$ be its children in the join tree, and assume that $f(m_1) > f(m_2)$. Thus, $s$ should pair with $m_1$. In some specific cases, it is possible that the growth of $m_2$ will reach saddle $s$ before $m_1$. This can lead to problems with persistence pairing. This situation can occur if, for example, $m_1$ is first grown to reach some other saddle $s'$; then, $m_2$ is grown to reach $s$ and pairs with $s$. Finally, due to a false case, $m_1$ must be re-grown, and when it is re-grown, its growth now terminates at $s$.

We handle this situation as follows. Let $s$ be an interior saddle point that has already been paired with points $\{m_i\}$. Let $m$ be a new point whose growth just terminated at $s$. If, for some $i$, we have that $f(m_i) < f(m)$, then an out-of-order growth occurs. To resolve this issue, we treat it as a false positive. Similarly, if the growth of $m$ terminates at $s$, but $s$ has already been grown as well, then this also signifies an out-of-order growth. We likewise treat this case as a false positive.

\para{Simultaneous false positive and false negative.} 
Suppose that when growing a minimum $m$, we discover that edge $ms \in T'_\varepsilon$ but $ms \notin T_\varepsilon$. As a result, $ms$ represents a false positive. Further, suppose that there exists some saddle $s'$ such that $ms' \in T_\varepsilon$. When this occurs, $ms'$ is a false negative. Therefore, $m$ is associated with both a false positive and a false negative. Let $R_p$ be the set of points associated with false positive $ms$. Let $R_n$ be the set of points associated with false negative $ms'$. In this case, we tighten the error bounds around all points in the region $R = R_p \cup R_n$.

When computing the average time per false case fixed in \cref{tab:time}, we treat this instance as a single false case (rather than two) as it is handled the same way as a single false case, except that $R$ is larger. In the `FC' column this case is marked as a false positive.

\subsection{Upper and Lower Bound Tightening}
\label{sec:tighten-bounds}

In this section, we give a detailed summary of the upper and lower error bound tightening process of TopoSZ, and compare it against our progressive bound tightening process.

\para{TopoSZ Bound Tightening.} 
For each false case that is detected, TopoSZ calculates a region $R$ of $\X$. Then, it tightens the upper and lower bounds $U(x)$ and $L(x)$ of points in $x \in R$. The region $R$ and how aggressively $L(x)$ and $U(x)$ are tightened depends on how many iterations are conducted to eliminate false cases. 
Let $n$ be the number of iterations (for Step 4 of TopoSZ), with $n=1$ initially. Let $f$ represent the input (ground truth) scalar field, and let $f'$ represent the decompressed scalar field. 
Let $T$ and $T'$ respectively represent their contour trees. 
Let $T_\varepsilon$ and $T'_\varepsilon$ respectively represent the  contour trees after persistence simplification with a threshold $\varepsilon$.

First, a region $R$ is calculated. Define an $m$-layer neighborhood of a point $x$ as the set of all points $y$ such $||x-y||_\infty \leq m$. In the event of a false positive edge $e'$ that is present in $T'_\varepsilon$ but not $T_\varepsilon$, $R$ is initially set equal to the region corresponding to $e'$ in the segmentation induced by $T'_\varepsilon$. Let $s'$ be the saddle point that is a vertex of $e'$. An $n$-layer neighborhood surrounding $s'$ is added to $R$.

In the event of a false negative edge $e$ that is in $T_\varepsilon$ but absent in $T'_\varepsilon$, $R$ is initially set equal to the region corresponding to $e$ in the segmentation induced by $T'_\varepsilon$. $R$ is then expanded by adding an $n$-layer neighborhood of $R$. TopoSZ handles false types in the same way that it handles false negatives.

Then, TopoSZ tightens the $L(x)$ and $U(x)$ bounds around points $x \in R$. Let $k_0 = \min\{f(x) : x \in R\}$ and $k_{n+1} = \max\{ f(x) : x \in R \}$. TopoSZ calculates $n+1$ intervals $[k_0, k_1], [k_1,k_2], \ldots, [k_n,k_{n+1}]~\subset~\R$ such that, for each interval $I$, approximately $\frac{1}{n+1}$ of points $x \in R$ satisfy $f(x) \in I$. For each point $x \in R$, if $f(x) \in [k_i, k_{i+1}]$, then $L(x)$ and $U(x)$ are adjusted according to $L(x) \leftarrow \max(L(x), k_i)$ and $U(x) \leftarrow \min( U(x), k_{i+1})$.

\para{Difference with our framework.} 
In our progressive error tightening procedure, we compute the regions $R$ in mostly the same manner as TopoSZ, but with changes. First, because we work with merge trees, we use the merge-tree-induced segmentation, rather than the contour-tree-induced segmentation. 

Second, in the case of a false positive, we do not grow the region around the saddle for the first three iterations; and in the case of a false negative, we do not grow the region for the first three iterations. We have found that this strategy leads to less points with tighter error bounds. On the other hand, we divide $R$ into $2^n$ intervals, rather than $n+1$, as we have found that this leads to faster convergence. Additionally, if a false positive persists for at least six iterations, we instead handle it the same way as a false negative. We have found that this also leads to faster convergence.
\section{Compressor Configurations and Reconstruction Quality Curves}
\label{sec:reconstruction-quality-extra}

The TTK implementation of TopoQZ is tightly coupled with ZFP, which relies on two parameters: the persistence threshold $e$ and the pointwise error bound $\zeta$ associated with ZFP. 
This gives rise to a total pointwise error upper-bounded by $e+\zeta$. Admittedly, this bound may not be tight and we estimate the pointwise error bound of TopoQZ to be lower-bounded by $\max\{e, \zeta\}$ and upper-bounded by $e+\zeta$. Decreasing this bound will generally improve the run time of TopoQZ.   

\begin{figure}[!ht]
\centering
\includegraphics[width=0.8\columnwidth]{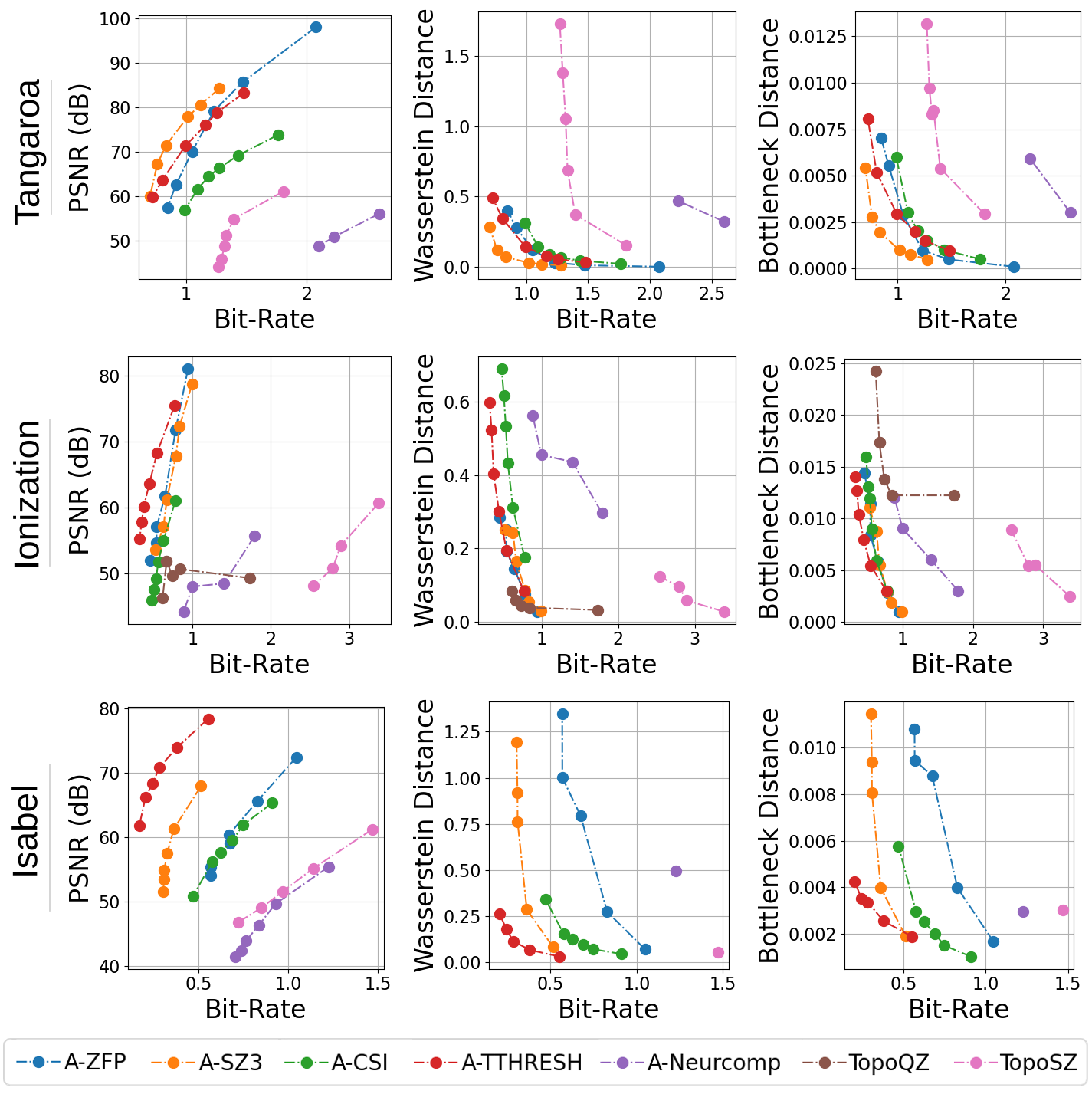}
\vspace{-4mm}
\caption{PSNR, bottleneck distance, and Wasserstein distance versus bit-rate for each compressor for the Tangaroa, Ionization, and Isabel datasets with $\varepsilon = 0.04$ ($e = 0.04$ for TopoQZ).}
\label{fig:reconstruction-quality-appendix}
\vspace{-2mm}
\end{figure}

We create plots demonstrating the trade off between bit-rate and respectively PSNR, Wasserstein distance $d_W$ and bottleneck distance $d_B$ for each augmented compressor, as well as TopoQZ and TopoSZ. 
Plots for Earthquake and QMCPACK are included in \cref{fig:reconstruction-quality}. Plots for the Tangaroa, Ionization, and Isabel datasets are included in \cref{fig:reconstruction-quality-appendix}. 
For the larger datasets, we are unable to compute the topological distances $d_W$ and $d_B$ using reasonable computational resources. 
To that end, for the Miranda, S3D, and SCALE-LETKF datasets, we only include plots of bit-rate vs PSNR, which can be found in \cref{fig:reconstruction-quality-appendix}. 
In the latter two large datasets (S3D and SCALE-LETKF), Augmented Neurcomp performs very poorly, and is an outlier. 
Thus, we do not plot its performance curves. 
For the Nyx dataset, because each individual trial has such a long run time, we do not generate any such curves. 
Some of the topological distances $d_W$ and $d_B$ could not be computed in reasonable time due to the large size or topological complexity of the dataset, thus are not plotted.

Most of the data is obtained for the augmented compressors and TopoSZ by setting $\varepsilon = 0.04$ and varying 
$\xi \in \{0.003, 0.006, 0.009, 0.012, 0.015, 0.018\}$. 
For TopoSZ, in order to obtain a wide range of values, we set $e = 0.04$ and vary $\zeta \in \{$ 0.003, 0.11, 0.22, 0.33, 0.44, 0.55 $\}$.

However, for roughly half of the curves associated with the augmented compressors, we use different values of $\xi$ due to a peculiarity that we observe. Under normal circumstances, if a compressor uses more bits to encode each data point (i.e., the bit-rate is higher), then it can more accurately reconstruct the data.
In some cases, however, we notice a negative correlation between bit-rate and reconstruction quality. This is unusual for compressors.

We do not know precisely why this trend sometimes occurs. We offer one possible explanation. For each $x \in \X$, let $U(x)$ and $L(x)$ denote the initial upper and lower bounds, respectively, for $x$, before error bound tightening. Recall that, before error bound tightening, it holds that $f(x) - \xi \leq L(X) \leq f'(x) \leq U(X) \leq f(x) + \xi$. Therefore, as $\xi$ decreases, the initial guess $f'(x)$ will become more accurate. Thus, as $\xi$ decreases, more of the contour tree will be preserved before the tightening step, and less tightening will occur. In practice, we notice that error bound tightening does not affect PSNR very much, but it does increase the bit-rate. As a result, if less error bound tightening occurs, then the bit-rate could be lower for lower values of $\xi$. We notice that when $\xi$ becomes very small, increasing $\xi$ always leads to increased bit-rate.

For each combination of a dataset and a compressor, we ensure that every time $\xi$ increases, the bit-rate also decreases. In order to do this, we sometimes need to vary $\xi$ for values other than the standard $0.003$ through $0.018$. To choose values for $\xi$, we typically search for new, lower values of $\xi$. However, for Augmented Neurcomp, due to its very long run times, we simply use  a subset of $\{0.003,0.006,0.009,0.012,0.015,0.018\}$ for which increasing $\xi$ always leads to decreasing bit-rate. We handle  TopoSZ and TopoQZ similarly to Augmented Neurcomp when the run times are long. The values of $\xi$ used to generate the plots are shown in \cref{tab:odd-parameter-configurations}.

\begin{figure}[!ht]
\centering
\includegraphics[width=0.8\columnwidth]{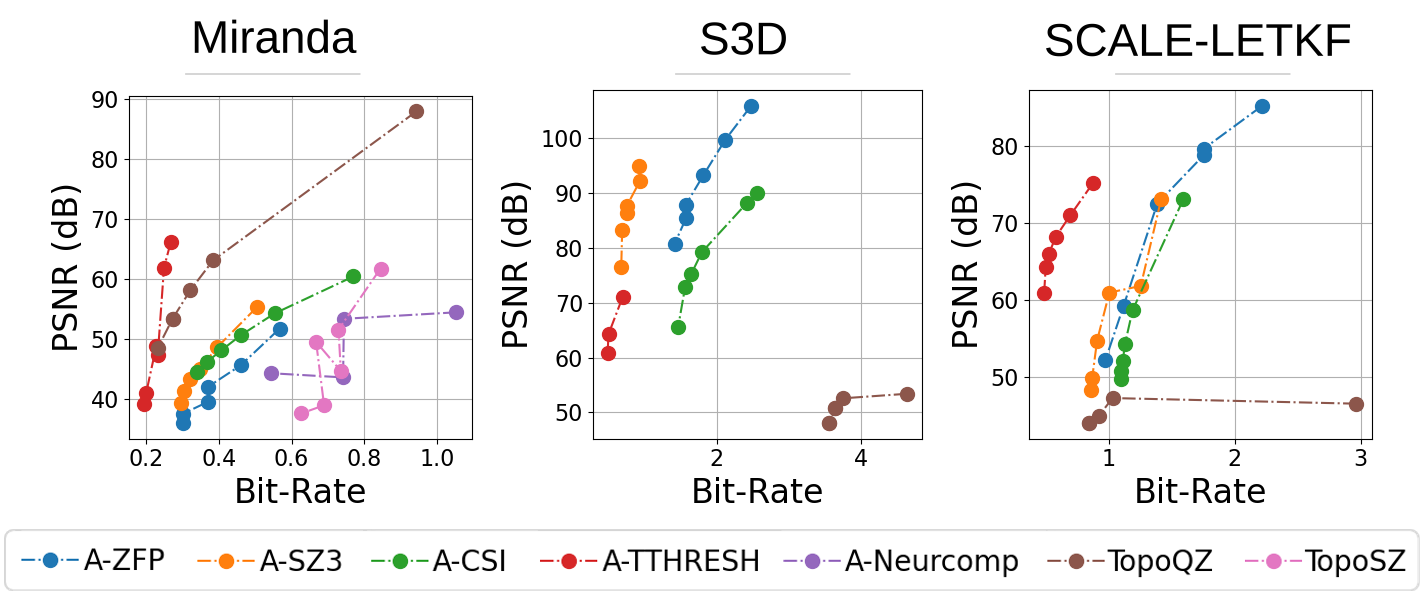}
\vspace{-4mm}
\caption{PSNR versus bit-rate for each compressor for the Miranda, S3D, and SCALE-LETKF datasets with $\varepsilon = 0.04$ and $e = 0.04$ for TopoQZ. Augmented Neurcomp is an outlier for S3D and SCALE-LETKF (bit-rate $\approx 9$) and is omitted.}
\label{fig:reconstruction-quality-no-topology}
%\vspace{-2mm}
\end{figure}

\begin{table}[!ht]
\vspace{2mm}
\resizebox{\columnwidth}{!}{\begin{tabular}{ll|l}
\hline
Dataset                                                  & Compressor                                                & Error Bounds                                                                                                                 \\ \hline
QMCPACK                                                  & SZ3                                                       & 0.0006, 0.00108, 0.00156, 0.00204, 0.00252                                                                                   \\
QMCPACK                                                  & TTHRESH                                                   & 0.0005, 0.0006, 0.0007, 0.0008, 0.0009, 0.001                                                                                \\
QMCPACK                                                  & TopoSZ                                                    & 0.002, 0.003, 0.006, 0.009, 0.012, 0.015                                                                                     \\ \hline
Tangaroa                                                 & ZFP                                                       & 0.0001, 0.0005, 0.001, 0.003, 0.006, 0.009                                                                                   \\
Tangaroa                                                 & SZ3                                                       & 0.0005, 0.00075, 0.001, 0.002, 0.003, 0.006                                                                                  \\
Tangaroa                                                 & CSI                                                       & 0.0005, 0.001, 0.0015, 0.002, 0.003, 0.006                                                                                   \\
Tangaroa                                                 & TTHRESH                                                   & 0.001, 0.0015, 0.002, 0.003, 0.006, 0.009                                                                                    \\
Tangaroa                                                 & Neurcomp                                                  & 0.003, 0.006, 0.009                                                                                                          \\ \hline
Earthquake                                               & SZ3                                                       & 0.001, 0.002, 0.003, 0.006, 0.009, 0.012                                                                                     \\
Earthquake                                               & CSI                                                       & 0.001, 0.002, 0.003, 0.006, 0.009, 0.012                                                                                     \\
Earthquake                                               & TTHRESH                                                   & 0.001, 0.002, 0.003, 0.006, 0.009, 0.012                                                                                     \\
Earthquake                                               & Neurcomp                                                  & 0.003, 0.006                                                                                                                 \\ \hline
Ionization                                               & ZFP                                                       & 0.001, 0.003, 0.006, 0.009, 0.012, 0.015                                                                                     \\
Ionization                                               & SZ3                                                       & 0.001, 0.002, 0.003, 0.006, 0.009, 0.012                                                                                     \\
Ionization                                               & Neurcomp                                                  & 0.003, 0.006, 0.009, 0.012                                                                                                   \\
Ionization                                               & TopoSZ                                                    & 0.003, 0.006, 0.009, 0.012                                                                                                   \\ \hline
Isabel                                                   & CSI                                                       & 0.001, 0.0015, 0.002, 0.0025, 0.003, 0.006                                                                                   \\
\begin{tabular}[c]{@{}l@{}}Isabel\\ $\quad$\end{tabular} & \begin{tabular}[c]{@{}l@{}}TTHRESH\\ $\quad$\end{tabular} & \begin{tabular}[c]{@{}l@{}}0.002, 0.003, 0.004, 0.005, 0.006, 0.009;\\ Only 0.003 is shown for $d_W$ and $d_B$.\end{tabular} \\
Isabel                                                   & TopoSZ                                                    & Only 0.003 is shown for $d_W$ and $d_B$                                                                                      \\ \hline
Miranda                                                  & TTHRESH                                                   & 0.002, 0.003, 0.006, 0.009, 0.012, 0.015                                                                                     \\ \hline
S3D                                                      & ZFP                                                       & 0.00005, 0.0001, 0.0002, 0.0003, 00005, 0.0007                                                                               \\
S3D                                                      & SZ3                                                       & 0.0001, 0.00025, 0.0003, 0.0004, 0.0005, 0.001                                                                               \\
S3D                                                      & CSI                                                       & 0.00008, 0.0001, 0.0003, 0.0005, 0.0007, 0.002                                                                               \\
S3D                                                      & TTHRESH                                                   & 0.003, 0.006, 0.009                                                                                                          \\ \hline
SCALE-LETKF                                              & ZFP                                                       & 0.0005, 0.0008, 0.001, 0.002, 0.003, 0.006                                                                                   \\
SCALE-LETKF                                              & SZ3                                                       & 0.001, 0.002, 0.003, 0.006, 0.009, 0.012                                                                                     \\
SCALE-LETKF                                              & CSI                                                       & 0.001, 0.003, 0.006, 0.009, 0.012, 0.015                                                                                     \\
SCALE-LETKF                                              & TTHRESH                                                   & 0.002, 0.003, 0.004, 0.005, 0.006, 0009                                                                                      \\ \hline
\end{tabular}}
\vspace{-3mm}
\caption{Parameter configurations used to generate \cref{fig:reconstruction-quality}, \cref{fig:reconstruction-quality-appendix}, and \cref{fig:reconstruction-quality-no-topology} if they are different from the default values.}
\label{tab:odd-parameter-configurations}
\end{table}
\section{Additional Experiments: Parameter Variations and Ablation Study}
\label{sec:other-experiments}

We describe additional experimental results with parameter variations and ablation studies. 
We describe the results by varying $\varepsilon$ and $\xi$ in~\cref{sec:vary-epsilon} and \cref{sec:vary-xi} respectively. We describe ablation studies for logarithmic-scaling quantization and progressive error bound tightening in~\cref{sec:ablation}. We only include results for the Earthquake and Ionization datasets due to space constraints. The trends reported are consistent across all datasets tested. In all plots of compression and decompression times, we omit the results of augmented Neurcomp as it is much slower than the other augmented compressors.

\subsection{Experiments with Varying Persistence Threshold}
\label{sec:vary-epsilon}

\begin{figure*}[!ht]
\centering
\includegraphics[width=\linewidth]{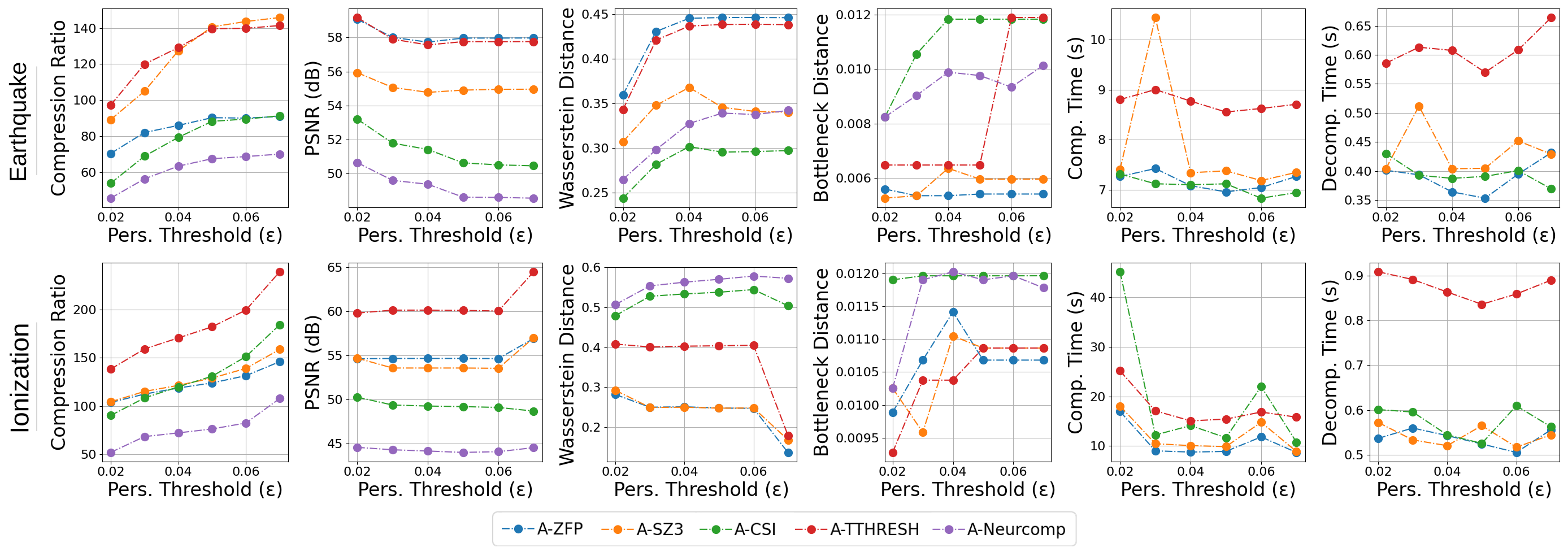}
\vspace{-6mm}
\caption{Each of the six evaluation metrics measured for the Earthquake and Ionization datasets as $\varepsilon$ varies. We fix $\xi = 0.012$. The augmented compressors are given as A-ZFP, A-SZ3, etc.}
\label{fig:epsilon-table}
\end{figure*}

We measure the effect of varying the persistence threshold $\varepsilon$ on each of our six evaluation metrics for each augmented compressor on each dataset (except Nyx). We fix $\xi = 0.012$ and vary $\varepsilon \in \{0.02,0.03,0.04,0.05,0.06,0.07\}$. The results for the Earthquake and Ionization datasets are shown in~\cref{fig:epsilon-table}.

We observe that there is a clear positive correlation between $\varepsilon$ and compression ratio. This trend is reasonable: when $\varepsilon$ is larger, the simplified contour tree becomes simpler, and thus less precision is required to preserve the contour tree, leading to a high compression ratio. There is no clear correlation between $\varepsilon$ and any other evaluation metrics.

\subsection{Experiments with Varying Pointwise Error Bound}
\label{sec:vary-xi}

We measure the effect of varying the pointwise error bound $\xi$ on each of our six evaluation metrics for each augmented compressor on four datasets (QMCPACK, Tangaroa, Earthquake, and Ionization). We fix $\varepsilon = 0.04$ and vary $\xi$ using the same values that were used to measure the trade-off between reconstruction quality and compression ratio. See \cref{sec:reconstruction-quality-extra} for specific parameter values.

\begin{figure*}[]
    \includegraphics[width=\linewidth]{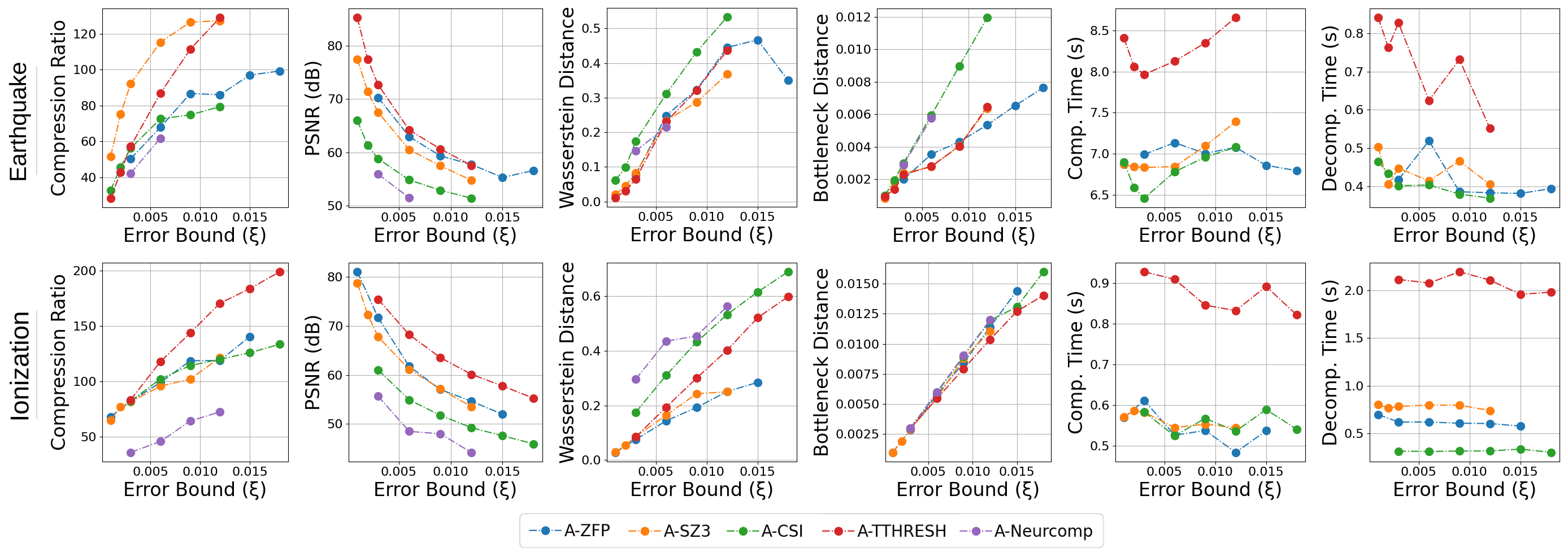}
    \vspace{-6mm}
    \captionof{figure}{Each of the six evaluation metrics measured for the Earthquake and Ionization datasets for each lossy compressor as $\xi$ varies. We fix $\varepsilon = 0.04$. The augmented compressors are given as A-ZFP, A-SZ3, etc.}
    \label{fig:xi-table}    
\end{figure*}

The results are shown in \cref{fig:xi-table}. There is a clear positive correlation between $\xi$ and compression ratio. This trend is logical, as less information should need to be stored in order to maintain a higher error bound. Likewise, there is a clear negative trend between $\xi$ and reconstruction quality in terms of each PSNR, Wasserstein distance, and bottleneck distance. This trend is also logical as a higher error bound means that the reconstructed data does not need to be as faithful to the ground truth data. There is no clear trend between $\xi$ and compression or decompression time.

\subsection{Ablation Study}
\label{sec:ablation}

In our framework, there are two main components: logarithmic-scaling quantization and progressive bound tightening. We perform an ablation study to separately justify the benefits of these two components, in comparison with TopoSZ, which utilizes linear-scaling quantization and iterative bound tightening. 
We independently analyze each of the six evaluation metrics across five datasets, for all augmented compressors. 

In addition to studying iterative bound tightening based on the contour tree (as is used by TopoSZ), we also implement an iterative bound tightening based on the merge tree for the ablation study, where we iteratively tighten the upper and lower bounds until there are no false cases in the join and split trees respectively. For linear-scaling quantization, we use quantization intervals of width $\xi$ (like in TopoSZ), rather than the standard width of $2\xi$.

\begin{figure*}[!ht]
    \includegraphics[width=\linewidth]{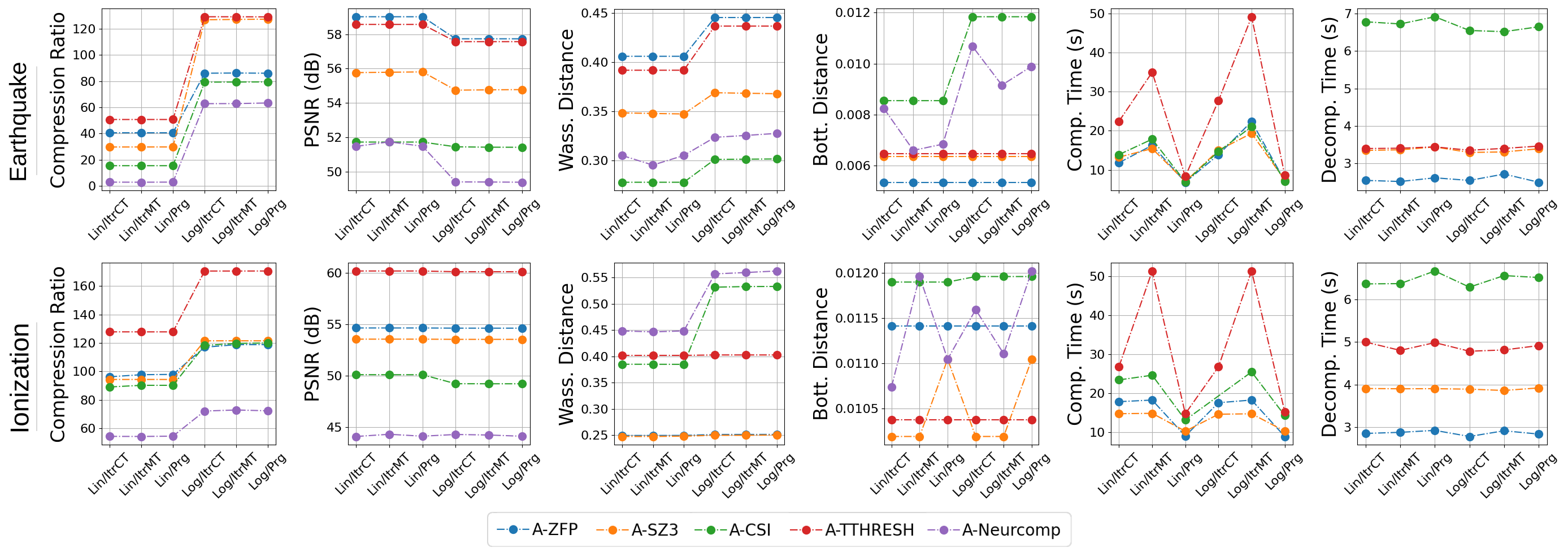}
    \vspace{-6mm}
    \captionof{figure}{Each of the six evaluation metrics measured for the Earthquake and Ionization datasets for each lossy compressor in an ablation study. Augmented compressors are given as A-ZFP, A-SZ3, etc. Run times from augmented Neurcomp, and those from augmented CSI are excluded as they are much higher compared to the others. Each point on the $x$-axis represents a combination of one quantization strategy (linear-scaling vs logarithmic-scaling) and one tightening strategy: iterative with contour trees such as in TopoSZ; iterative with merge trees; or progressive bound tightening.
    Lin: linear scaling quantization. 
    Log: logarithmic-scaling quantization. 
    Prg: progressive upper and lower bound tightening. 
    ItrCT: contour-tree based iterative upper and lower bound tightening. 
    IterMT: merge-tree based iterative upper and lower bound tightening.}
    \label{fig:ablation-table}
\end{figure*}

The results of the ablation study are shown in~\cref{fig:ablation-table}. The augmented compressors are given by A-ZFP, A-SZ3, etc. In each line chart, columns whose label start with `Lin' use linear-scaling quantization, whereas columns whose label starts with `Log' use logarithmic-scaling quantization. Columns whose labels end with `ItrCT,' `ItrMT,' and `Prog' refer to trials that use contour-tree-based iterative tightening, merge-tree-based iterative tightening, and progressive tightening respectively. For compression times, we omit the time collected for Ionization dataset using CSI and Log/ItrCT as it was much higher than the times reported by other compressors and configurations.

In \cref{fig:ablation-table}, we can see that logarithmic-scaling quantization always produces a higher compression ratio than linear-scaling quantization. Progressive or iterative tightening do not noticeably affect the compression ratio. The PSNR is essentially constant across each trial, although logarithmic-scaling quantization typically produces slightly lower PSNR values than linear-scaling quantization. The Wasserstein distance is similar across trials, but sometimes we observe slight increases to Wasserstein distance when using logarithmic-scaling quantization rather than linear-scaling quantization. We do see large fluctuations in the bottleneck distance between trials, and sometimes logarithmic-scaling quantization leads to an increase. However, across all trials there is no clear pattern.

In terms of compression time, we can see that progressive bound tightening outperforms iterative bound tightening in every trial. Also, it appears that iterative tightening based on the contour tree outperforms iterative tightening based on the merge trees in most trials. This trend suggests that the primary benefit of tightening based on the join and split trees is that doing so enables progressive tightening strategies. When progressive tightening is used, using logarithmic-scaling quantization sometimes leads to slightly longer run times than linear-scaling quantization. Neither the tightening strategy or the quantization strategy appear to offer any advantage or disadvantage in terms of decompression time.

\section{Visual Artifacts Due to Lossy Compression}
\label{sec:visual-artifacts}

\begin{figure*}[]
\centering
  \begin{subfigure}[b]{0.02\textwidth}
    \raisebox{1.8\height}{\includegraphics[angle=90,width=\textwidth]{fig-colorBar}}
  \end{subfigure}
  \begin{subfigure}[b]{0.97\textwidth}
    \includegraphics[width=\linewidth]{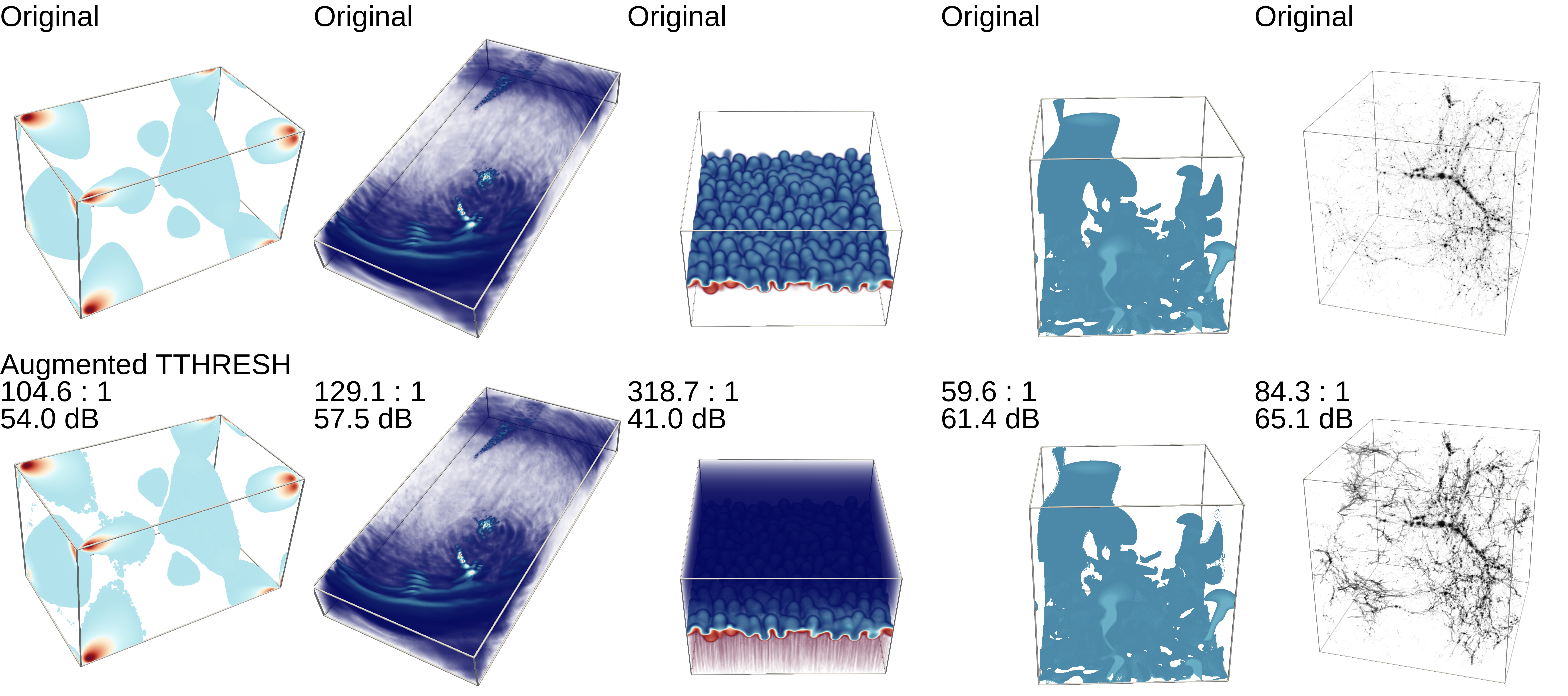}
  \end{subfigure}
\subfigsCaption{
Volume renderings of decompressed scientific datasets when compressed with augmented TTHRESH versus the input (ground truth), using transfer functions designed to produce the maximal visual difference between the two.
Top row: volume renderings of the input data. 
Bottom row: volume renderings of the decompressed data after compression with Augmented TTHRESH. 
Datasets from left to right: QMCPACK, Earthquake, Miranda, S3D, and Nyx.}
\label{fig:volume-render-adversarial}
\end{figure*}

Lossy compressors allow for some distortion of the data to achieve smaller compressed file sizes. Since we work with augmented lossy compressors, we might observe visual artifacts in the decompressed data due to the distortion introduced during the compression process. Some artifacts become  more visible depending on the choice of transfer functions during volume rendering. 

%When performing volume renderings of decompressed data after it is compressed with our framework, we notice some visual artifacts due to the selection of transfer functions. 

If the visualization is particularly sensitive to small changes in the transfer function, the decompressed data can appear lighter or darker than the input (ground truth) when visualized with the same transfer function. This typically occurs when one region of a volume has relatively uniform values. 

We include visualizations of the visual artifacts for five datasets for augmented TTHRESH in \cref{fig:volume-render-adversarial}. 
We cannot find volume renderings that produce noticeable artifacts for the remaining four datasets. 
We do not consider Neurcomp as its augmented model preforms the worst. The remaining augmented compressors yield similar artifacts. The Miranda dataset appears to contain the most visual artifacts, where a turbulent front divides two nearly uniform regions.
%-------------------------

\end{document}